\documentclass[pdflatex,sn-mathphys-ay]{sn-jnl}

\usepackage[T1]{fontenc}
\usepackage[utf8]{inputenc}
\usepackage{graphicx}
\usepackage{booktabs}
\usepackage{amsmath,amssymb}
\usepackage{amsthm}
\usepackage{listings}
\usepackage{xcolor}
\usepackage{array}
\usepackage{float}

\theoremstyle{thmstyletwo}
\newtheorem{example}{Example}

\newcounter{question}
\newlength{\prefixwidth}

\newcommand{\question}[2]{%
  \refstepcounter{question}%
  \def\prefix{\textbf{RQ\arabic{question}: }}%
  \settowidth{\prefixwidth}{\prefix}%
  \par\medskip
  \noindent
  \hangindent=\prefixwidth
  \hangafter=1
  \makebox[\prefixwidth][l]{\prefix\label{rq:#1}}#2\par\medskip
}

\newcommand{\questionRef}[1]{%
RQ\ref{rq:#1}%
}

\begin{document}

\title[Reducing the Cross-Model Tax]{Reducing the Cross-Model Tax: Query Optimization over Multi-Model Data}

\author[1]{\fnm{J\'achym} \sur{B\'art\'ik}}

\author[1]{\fnm{Filip} \sur{\v{S}trobl}}

\author*[1]{\fnm{Irena} \sur{Holubov\'a}}
\email{irena.holubova@matfyz.cuni.cz}

\affil*[1]{\orgdiv{Department of Software Engineering},
    \orgname{Faculty of Mathematics and Physics, Charles University},
    \orgaddress{\street{Malostransk\'e n\'am\v{e}st\'i 25},
    \city{Prague}, \postcode{118 00}, \country{Czech Republic}}}

\abstract{Querying across heterogeneous data models incurs overhead from query decomposition, result retrieval and conversion, and processing outside the underlying database systems. This paper investigates the extent to which, in a decomposition-based architecture, this cross-model tax results from decisions made by the unifying query processor rather than from heterogeneity alone.

We present a mapping- and capability-aware optimization approach that moves applicable processing into native query parts. It combines model-aware predicate pushdown, cross-model dependent joins, and non-redundant query-part construction within a unified pipeline spanning relational, document, and graph databases.

The approach is implemented in MM-quecat and evaluated using 20 read-only queries across PostgreSQL, MongoDB, and Neo4j, as well as a heterogeneous combination of the three systems in a single-machine, containerized deployment. For the query--environment combinations most affected by large intermediate results, predicate pushdown yields maximum observed latency reductions of up to two orders of magnitude and prevents the out-of-memory failures observed in the original single-DBMS experiments. Dependent execution further improves eligible external joins, while non-redundant construction reduces planning time for the largest evaluated graph plans, from hundreds of milliseconds to several milliseconds. The results show how established optimization principles can be applied across conceptual, mapping, data-model, and DBMS boundaries in decomposition-based multi-model query processing.}


\maketitle

\section{Introduction}
\label{sec:introduction}

Modern applications increasingly combine data with structurally distinct properties. Transactional records are naturally represented in relational tables, aggregate-oriented data in documents, and highly connected data in graphs. Multi-model data management addresses this heterogeneity by providing a common interface across multiple data models, either within one database system or through a unifying layer over specialized database management systems (DBMSs). Many widely used DBMSs have consequently extended their original models with additional representations~\citep{ranking}. Nevertheless, the practical adoption of multi-model data management has progressed faster than the development of equally mature foundations and query-processing methods for heterogeneous querying~\citep{10.1145/3323214,mm-taming,mm-look-forward}.

A unified interface simplifies access to heterogeneous data, but it does not remove differences between physical representations, query languages, and execution capabilities. A query spanning several models may have to be decomposed into model-specific parts, translated into native query languages, executed by different DBMSs, and completed by retrieving, converting, joining, and restructuring intermediate results in a unifying layer. This architecture resembles distributed and federated query processing, where performance depends on operator placement, the size of intermediate results that cross source boundaries, and native execution costs~\citep{opt-diverse-datasrc,opt-distributed}. We use the term \emph{cross-model tax} for the additional planning, retrieval, conversion, and unifying-layer processing overhead observed in such decomposition-based execution.

This overhead can determine whether a heterogeneous query remains practically executable. If a filter is evaluated only after data retrieval, records that will ultimately be discarded must still be returned, converted, and processed. If the inputs of a cross-model join are retrieved independently, both may be materialized even when values from one input could substantially restrict the other. During planning, different orders in which the same mandatory physical patterns are added may also yield equivalent query parts and unnecessarily enlarge the search space. The principal bottleneck can therefore lie not in native query execution itself, but in how the unifying processor decomposes, coordinates, and completes the query.

Existing development of multi-model query processors, including MM-quecat, has primarily established functional expressiveness: which heterogeneous queries can be represented, translated, and evaluated correctly. Functional coverage alone does not ensure practical performance. A processor may support the required constructs while retrieving unnecessarily large results, exhausting available memory, or repeatedly constructing equivalent plans. Efficient multi-model querying, therefore, also requires optimization mechanisms that relate conceptual query operators to heterogeneous physical mappings and determine which operations can be delegated to the underlying systems.

Established database optimization principles provide a starting point. Predicate pushdown reduces intermediate results by evaluating filters closer to stored data~\citep{opt-diverse-datasrc,opt-distributed}. Dependent joins use values obtained from one query branch to restrict another before retrieval; related variants are known as bind joins, D-joins, lateral joins, or tuple-substitution joins~\citep{opt-diverse-datasrc,opt-orthogonal}. Avoiding equivalent construction paths similarly reduces planning overhead. Applying these principles in a general multi-model processor is not, however, a purely algebraic task. The optimizer operates over a representation independent of the physical data models, must establish which conceptual values each mapping provides, and must account for the different operations expressible by the target DBMSs.

This paper presents a mapping- and capability-aware optimization approach for decomposition-based multi-model query processing. It addresses three sources of avoidable overhead: late predicate evaluation, independent retrieval of cross-model join inputs, and redundant exploration during query-part construction. The approach moves applicable predicates into native query parts, introduces execution dependencies between plan branches, and adds mandatory kind patterns without exploring their permutations. Lightweight query-driven cardinality estimates support dependent-join orientation but are not intended as a complete heterogeneous cost model.

The contribution is not predicate pushdown, dependent execution, or search-space reduction in isolation. It is their realization across four boundaries: conceptual query semantics, physical mapping coverage, data-model-specific representation, and native DBMS capabilities. Unlike an optimizer that starts from source-specific relations or an already assigned physical plan, the considered processor starts from operators over a unified conceptual schema. It must first determine which mappings provide the required conceptual values and then whether the corresponding native query builders can execute an operation without changing its semantics. A dependent restriction may consequently originate in one physical model, propagate through model-independent plan operators, and be consumed by a query part translated into another native language. Query-part construction must likewise retain genuinely different mapping alternatives while eliminating only provably equivalent construction paths.

We implement the approach in MM-quecat~\citep{mmquecat}, the query-processing component of the category-theory-based MM-cat framework~\citep{mmcat}. MM-quecat evaluates queries expressed in the Multi-Model Query Language (MMQL) over a unified conceptual schema and mappings to relational, document, and graph databases. Its architecture separates conceptual query processing, mapping selection, native query generation, execution by the underlying DBMSs, and processing retained in the unifying layer. The proposed techniques are implemented for PostgreSQL, MongoDB, and Neo4j and are evaluated in relational, document, graph, and heterogeneous environments.

The contributions of this paper are as follows:
\begin{enumerate}
    \item We formulate a mapping- and capability-aware optimization approach that links transformations over conceptual operators to physical mapping coverage, native execution capabilities, and operations retained in the unifying layer.
    \item Within this approach, we realize model-aware predicate pushdown over relational, document, and graph query parts; dependent joins whose restrictions can propagate through heterogeneous plan subtrees; and non-redundant construction that eliminates equivalent insertion orders of mandatory kind patterns while preserving distinct mapping alternatives.
    \item We implement the complete pipeline in MM-quecat and evaluate its incremental effects across PostgreSQL, MongoDB, and Neo4j, as well as a heterogeneous combination of the three systems. The evaluation identifies the plan and mapping properties under which the techniques reduce execution or planning overhead and characterizes the remaining bottlenecks.
\end{enumerate}

In the evaluated architecture and workload, the maximum observed latency reduction after predicate pushdown reaches two orders of magnitude, and eligible dependent joins provide further improvements of up to one order of magnitude. Predicate pushdown also prevents the out-of-memory failures observed in the original single-DBMS experiments. Non-redundant construction reduces planning time for the largest evaluated graph plans from almost 600 milliseconds to several milliseconds. These maxima characterize particular query--environment combinations rather than uniform improvements across the workload. Together, the results show that a unifying layer can reduce avoidable overhead by coordinating mappings and native capabilities and retaining only operations that cannot be delegated safely.

The remainder of the paper is organized as follows. Section~\ref{sec:preliminaries} introduces the multi-model query-processing model. Section~\ref{sec:optimization} presents the sources of cross-model overhead and the proposed optimization methods. Section~\ref{sec:implementation} describes their implementation in MM-quecat. Section~\ref{sec:evaluation} reports the experimental evaluation and discusses threats to validity. Section~\ref{sec:related-work} reviews related work, and Section~\ref{sec:conclusion} concludes the paper.
\section{Preliminaries and Query-Processing Model}
\label{sec:preliminaries}

This section defines the query-processing abstractions used in the remainder of the paper. It first summarizes the differences between relational, document, and graph representations that affect operator placement. It then introduces the unified schema, physical mappings, query language, and decomposition-based execution model of MM-quecat.

\subsection{Heterogeneous Data Models}
\label{sec:data-models}

The relational, document, and graph data models differ in how they represent records, relationships, and complex values. These differences affect both schema design and the set of operations that can be delegated efficiently to the corresponding database management systems~\citep{mm-taming,mm-look-forward}.

The \textit{relational model} organizes records into relations defined over fixed sets of attributes. Relational schemas commonly reduce redundancy through normalization, which improves data integrity but may distribute related data across multiple relations and, consequently, require joins during query execution. PostgreSQL~\citep{postgresql} is used as the relational DBMS in our implementation.

The \textit{document model} stores hierarchically structured records that may contain nested objects and variable-size arrays. Related data can therefore be embedded in a single document rather than distributed across several relations. Document schemas may intentionally introduce redundancy to reduce the need for joins, a process commonly referred to as denormalization~\citep{denormalization}. MongoDB~\citep{mongodb} is used as the document DBMS in our implementation.

The \textit{graph model} represents records as vertices and edges and is well-suited to data where relationships and traversal paths are central to the workload. Neo4j~\citep{neo4j} is used as the graph DBMS in our implementation, with Cypher~\citep{cypher-manual} as its native declarative query language.

\begin{figure}[ht]
    \centering
    \includegraphics[width=\textwidth]{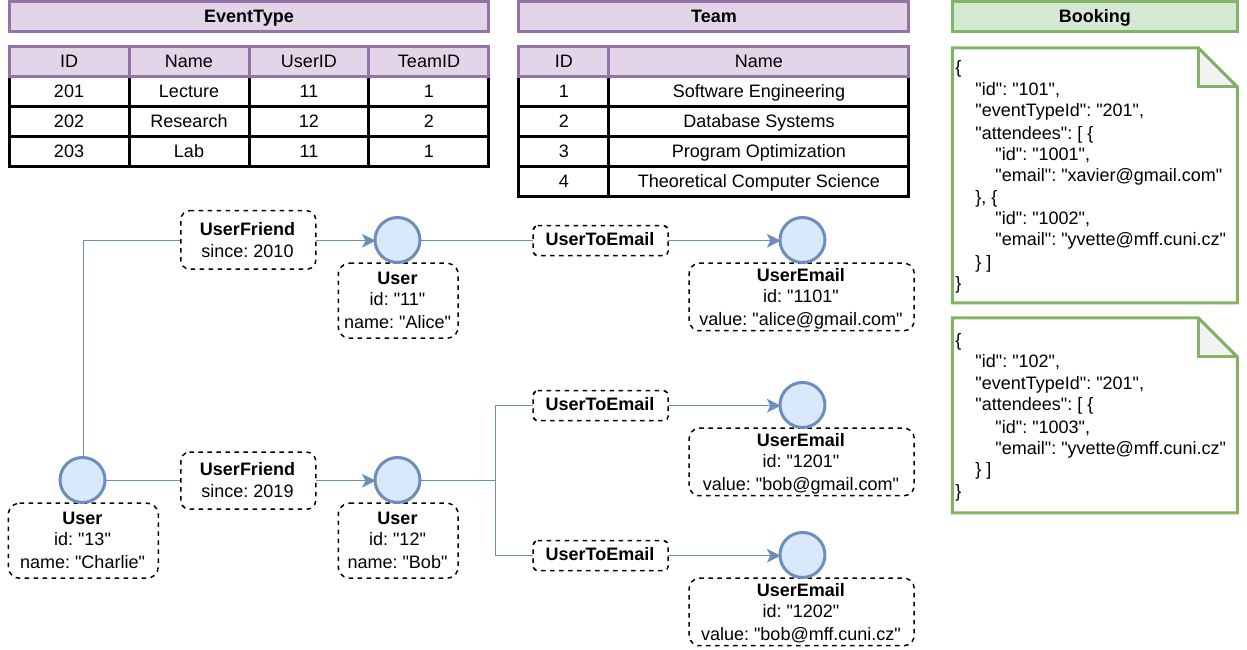}
    \caption{An example of multi-model data combining relational, document, and graph representations.}
    \label{fig:example-data}
\end{figure}

\begin{example}
Figure~\ref{fig:example-data} combines the three representations. Event types and teams are stored in tables (purple); bookings and attendees are stored in documents (green); and users and their relationships are stored in a graph (blue). References connect these representations across model boundaries: booking documents refer to relational event types, while event types refer to users represented in the graph database. Queries following these references must therefore combine data from PostgreSQL, MongoDB, and Neo4j, making operator placement and intermediate-result size important for performance.
\qed
\end{example}

A \textit{multi-model DBMS} may support several models within a single system, while a multi-model processing layer may provide a unified interface over several specialized systems. Many originally single-model systems have added further representations, whereas systems such as OrientDB~\citep{orientdb} were designed as multi-model systems from their initial release. Despite this practical development, multi-model data management still involves open theoretical and implementation challenges~\citep{mm-taming}. Relational algebra alone does not directly capture all structures of the models under consideration, which has also motivated model-independent and category-theory-based approaches.

The optimization challenge is therefore not only that the models represent data differently, but also that their DBMSs provide different query capabilities. An operation that can be evaluated locally in one system may require additional processing in another. A multi-model query processor must reason jointly about the conceptual operation, the physical representation, and the capabilities of the selected DBMS.

\subsection{Unified Schema and Mappings}
\label{sec:schema-mappings}

We implement and evaluate the proposed techniques in MM-quecat, the query-processing component of the MM-cat framework~\citep{mmcat,mmquecat}. MM-cat separates a unified conceptual representation from physical representations managed by the underlying DBMSs. A query is therefore stated without selecting a data model or DBMS, but its execution must subsequently be grounded in mappings to concrete physical data.

MM-cat uses category theory as a model-independent foundation. In MM-cat, a \emph{schema category} represents the unified conceptual schema. Schema objects correspond to entities and attributes, schema morphisms represent relationships, and composite morphisms describe paths consisting of several relationships. The complete formal definition is given in the original description of MM-cat~\citep{mmcat}.

\begin{figure}[ht]
    \centering
    \includegraphics[width=\textwidth]{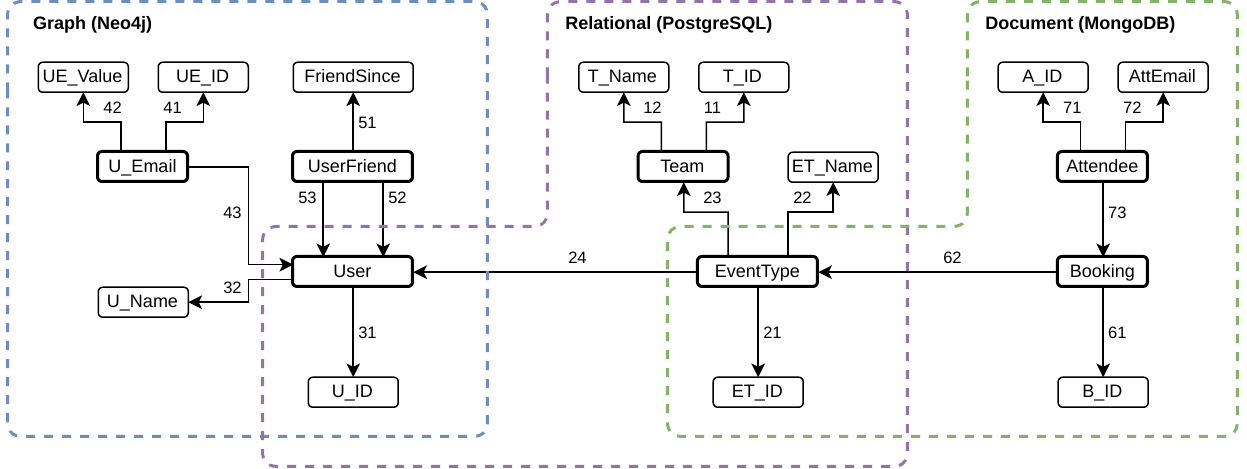}
    \caption{A schema category covering relational, document, and graph representations.}
    \label{fig:example-schema}
\end{figure}

\begin{example}
Figure~\ref{fig:example-schema} shows the unified schema category corresponding to the physical data in Figure~\ref{fig:example-data}. Boxes represent schema objects, solid connecting lines represent morphisms, and the numerical labels identify morphism signatures. The blue-dashed region on the left marks the schema fragment represented in Neo4j, the purple-dashed region in the center marks the fragment represented in PostgreSQL, and the green-dashed region on the right marks the fragment represented in MongoDB. The regions overlap at conceptual objects that span physical representations, showing how a single unified schema can cover data stored in different models and DBMSs.
\qed
\end{example}

The unified schema is connected to physical data through \emph{mappings}. A mapping associates a conceptual schema fragment with one \emph{kind}, that is, a collection of records managed by an underlying DBMS. Each mapping has a root schema object and a tree of properties reachable through schema morphisms. The physical data model determines whether these properties are represented as scalar values, arrays, references, or nested structures.

The same conceptual fragment may therefore have different physical representations and expose different opportunities for native filtering, joining, and projection. PostgreSQL mappings contain relational records without nested properties, MongoDB mappings may contain nested documents and arrays, and Neo4j mappings may contain arrays but not nested property structures.

\begin{example}
The \emph{Booking} object in the green MongoDB region of Figure~\ref{fig:example-schema} is mapped to the green booking documents shown on the right of Figure~\ref{fig:example-data}. Each document contains the booking identifier, a reference to an event type, and an array of embedded attendee records. The mapping, therefore, covers not only the \emph{Booking} object itself, but also the identifier, event-type reference, and attendee-related properties reachable from it. A query expressed over the unified schema does not explicitly select this MongoDB representation; the processor must identify the mapping and determine which of its properties are required.
\qed
\end{example}

Table~\ref{tab:query-processing-terms} summarizes the main terms used throughout the paper.

\begin{table}[ht]
    \centering
    \scriptsize
    \caption{Main concepts of the decomposition-based query-processing model.}
    \label{tab:query-processing-terms}
    \begin{tabular}{lp{0.76\textwidth}}
        \toprule
        \textbf{Term} & \textbf{Meaning} \\
        \midrule
        Mapping & A correspondence between a conceptual schema fragment and one physical kind. \\
        Kind & A collection of records managed by a particular underlying DBMS. \\
        Kind pattern & The part of a mapping required to cover selected properties of a query. \\
        Query part & One or more compatible kind patterns translated into a single native query. \\
        A datasource node & A QEP node that executes a query part in an underlying DBMS. \\
        Unifying layer & The model-independent layer that coordinates native queries and evaluates operations that cannot be delegated. \\
        \bottomrule
    \end{tabular}
\end{table}

\subsection{Multi-Model Query Language}
\label{sec:mmql}

MM-quecat accepts queries expressed in the Multi-Model Query Language (MMQL)~\citep{mmquecat}. MMQL allows users to query the unified schema without selecting physical mappings, data models, DBMSs, or their native query languages. Its design is primarily inspired by graph pattern languages, particularly SPARQL~\citep{sparql}.

An MMQL query contains a \texttt{WHERE} clause describing the requested schema fragment and a \texttt{SELECT} clause describing the structure of the result. The language also supports filters, value restrictions, optional patterns, set operations, aggregations, ordering, and result-size restrictions. The optimizations evaluated in this paper target the currently implemented select--project--join subset with filters; the broader language features are outside the experimental scope.

Patterns in the \texttt{WHERE} clause are expressed as triples of the form \emph{domain--morphism--codomain}. The domain and codomain may be variables or literals, while the morphism is identified by its signature. The operator \texttt{/} composes morphisms, and the unary operator \texttt{-} selects the corresponding dual morphism. Triples in the \texttt{SELECT} clause describe the hierarchical structure of the returned document.

Before physical planning, MM-quecat normalizes the query. Composite morphisms are decomposed into base morphisms connected through generated variables, dual morphisms are replaced by base morphisms with exchanged endpoints, and literal restrictions are represented uniformly. This allows later planning stages to operate over a common representation of conceptual patterns and filters.

\begin{example}
Listing~\ref{lst:mmql-example} shows an MMQL query over the schema in Figure~\ref{fig:example-schema}. It retrieves bookings, their identifiers, attendee email addresses, and the email addresses of the corresponding event-type hosts. The path expression \texttt{62/24/-43/42} traverses the conceptual path from a booking, through its event type and host user, to the user's email value. The expression \texttt{62/61 "201"} restricts the result to bookings associated with event type \texttt{201}, while the \texttt{FILTER} removes cases in which an attendee email equals a host email.
\qed
\end{example}

\begin{lstlisting}[
    basicstyle=\ttfamily\scriptsize,
    caption={Example MMQL query statement.},
    label={lst:mmql-example}
]
SELECT {
    ?booking
        id ?bId ;
        attendeeEmails ?attEmail ;
        hostEmails ?hostEmail .
}
WHERE {
    ?booking
        61 ?bId .
        -73/72 ?attEmail .
        62/24/-43/42 ?hostEmail .
        62/61 "201" .

    FILTER(?attEmail != ?hostEmail)
}
ORDER BY ?bId LIMIT 10
\end{lstlisting}

The implementation returns results as hierarchical documents and uses the same representation internally for intermediate results obtained from heterogeneous DBMSs. This common representation enables model-independent processing, but results must be retrieved through the respective drivers and converted before the remaining operations can be applied.

\subsection{Decomposition-Based Query Execution}
\label{sec:execution-model}

A \textit{query execution plan} (QEP) represents the concrete operations required to evaluate a query. It is commonly structured as a tree, with leaves accessing stored data and inner nodes processing the results of their children. A declarative query can usually be evaluated by several equivalent plans that produce the same result but may differ substantially in execution cost~\citep{opt-diverse-datasrc}.

In a conventional DBMS, the optimizer selects access paths, operators, and their order within one execution environment. A decomposition-based multi-model processor must additionally determine which physical mappings cover the conceptual query and where each operation is executed. Its QEP may therefore combine native queries evaluated by different DBMSs with model-independent operations over their intermediate results.

MM-quecat processes a query in four principal steps~\citep{mmquecat}:
\begin{enumerate}
    \item it normalizes the MMQL statement and decomposes it into conceptual query patterns;
    \item it identifies mappings covering these patterns and constructs candidate query parts and QEPs;
    \item it translates selected query parts into SQL, MongoDB aggregation pipelines, or Cypher and executes them in the underlying DBMSs;
    \item it converts the returned data into a common representation and applies the remaining operations in the unifying layer.
\end{enumerate}

This decomposition creates the optimization space addressed in this paper. A \emph{kind pattern} selects the properties of one mapping required by the query, while a \emph{query part} combines compatible kind patterns that can be translated into one native query. Query parts are represented by datasource nodes in the QEP. Neighboring patterns and operations may be merged into a larger native query part only when they belong to the same execution environment, the selected mappings provide all required values, and the corresponding query builder supports the complete operation. Otherwise, the operation remains in the unifying layer.

Operator placement directly affects the cross-model tax. A filter evaluated by an underlying DBMS reduces the data returned to the unifying layer, whereas the same filter evaluated after retrieval requires additional conversion and processing. Similarly, independently retrieving both inputs of a join may be more expensive than using one input to restrict the other. Query-part construction may itself become expensive if the planner repeatedly generates equivalent alternatives.

The cost of decomposition-based execution can therefore be divided into four connected components:
\begin{enumerate}
    \item construction and selection of the multi-model QEP;
    \item execution of native queries by the underlying DBMSs;
    \item retrieval and conversion of intermediate results across native query boundaries;
    \item evaluation of the remaining operations in the unifying layer.
\end{enumerate}

The proposed techniques address three corresponding sources of avoidable overhead: late predicate evaluation, independent retrieval of cross-model join inputs, and redundant construction of query parts. Predicate pushdown moves applicable filters into native query parts, dependent joins introduce information flow between plan branches, and non-redundant construction removes equivalent planning paths while preserving distinct physical alternatives.
\section{Cross-Model Query Optimization}
\label{sec:optimization}

This section presents three techniques for reducing the cross-model tax:

\begin{itemize}
    \item predicate pushdown reduces the data retrieved from underlying database management systems (DBMSs);
    \item dependent joins reduce the inputs to joins executed in the unifying layer; and
    \item non-redundant query-part construction reduces planning overhead.
\end{itemize}

The techniques adapt established optimization principles to decomposition-based multi-model query processing. Their contribution lies in their integration into one mapping- and capability-aware pipeline rather than in the individual principles themselves. Operators are first expressed over a unified conceptual schema, assigned to query parts covered by physical mappings, and finally translated into SQL, MongoDB aggregation pipelines, or Cypher. An optimization is applicable only when it preserves conceptual semantics, the selected mapping provides every required value, and the target query builder can express the complete operation.

\subsection{Sources of the Cross-Model Tax}
\label{sec:overhead-sources}

The execution model introduced in Section~\ref{sec:execution-model} has four principal cost components:
\begin{enumerate}
    \item construction and selection of a multi-model query execution plan;
    \item execution of native queries by the underlying DBMSs;
    \item retrieval and conversion of intermediate results across native query boundaries;
    \item processing of operations retained in the unifying layer.
\end{enumerate}

The first component depends mainly on the number of mappings, query parts, and candidate construction paths. The remaining components are strongly affected by operator placement. An operation executed by an underlying DBMS can use its native optimizer and may reduce the result before it crosses a native query boundary. The same operation, executed in the unifying layer, processes the data only after retrieval and conversion. Similar trade-offs are well known in distributed and heterogeneous query optimization, where execution cost depends not only on local operator cost, but also on data movement and the size of intermediate results crossing source boundaries~\citep{opt-diverse-datasrc,opt-distributed}. In this paper, the cross-model tax denotes these measurable planning, retrieval, conversion, and unifying-layer processing components in the evaluated architecture; it is not introduced as a universal cost metric.

Three sources of avoidable overhead are particularly important in the evaluated pipeline:

\begin{enumerate}
    \item Filters are initially placed above the query parts that provide their variables. Although this placement is correct, records that do not satisfy the filter may still be retrieved, transferred, converted, and possibly joined before they are discarded.
    \item When a join cannot be included in one native query, its inputs are retrieved independently and joined in the unifying layer. This may produce large intermediate results even when values from one input could substantially restrict the other.
    \item Query-part construction may explore several orders of adding the same mandatory kind patterns. These paths produce the same query part but may increase planning time factorially with the number of mandatory patterns.
\end{enumerate}

The following sections address these sources of overhead.

\subsection{Model-Aware Predicate Pushdown}
\label{sec:predicate-pushdown}

Predicate pushdown moves filters toward the leaves of a query execution plan so that they are evaluated before unnecessary intermediate results are retrieved. The principle is standard in relational and distributed query optimization because early filtering reduces both intermediate-result size and subsequent processing cost~\citep{opt-diverse-datasrc,opt-distributed}. In a multi-model processor, however, pushdown is not merely an algebraic rewrite. A filter is expressed over the unified conceptual schema and can be delegated only if the selected mapping provides every referenced value and the target query builder can represent the complete condition natively.

The optimization, therefore, consists of two stages. First, a filter is moved through model-independent plan operators while preserving query semantics. Second, when it reaches a native query part, its condition is translated into the query language of the corresponding DBMS.

Let \(f\) be a filter whose condition references a set of variables \(V_f\). The filter can be moved into a subtree only if the subtree provides every variable in \(V_f\) and the movement is valid for the parent operator.

\subsubsection{Transformation Rules}

Two consecutive filters can be reordered because evaluating predicates \(p_1\) and \(p_2\) is equivalent to evaluating their conjunction. A filter may therefore pass another filter while searching for a deeper valid position.

For an inner join with inputs \(A\) and \(B\), let \(V_A\) and \(V_B\) denote the sets of variables provided by the respective inputs. A filter can be moved into \(A\) when all variables referenced by the filter are provided by \(A\):
\[
V_f \subseteq V_A
\quad \Longrightarrow \quad
\sigma_f(A \Join B) \equiv \sigma_f(A) \Join B.
\]

The corresponding rule applies symmetrically to \(B\):
\[
V_f \subseteq V_B
\quad \Longrightarrow \quad
\sigma_f(A \Join B) \equiv A \Join \sigma_f(B).
\]

The rule does not generally apply to outer joins. For example, moving a predicate over the optional input of a left outer join may change whether unmatched records are preserved. The evaluated implementation, therefore, restricts this transformation to inner joins.

\subsubsection{Pushdown into Native Query Parts}

When a filter reaches a native query part, the processor checks whether the selected mapping provides all required properties and whether the corresponding query builder supports every operator used in the condition. If both conditions hold, the filter becomes part of the native SQL query, MongoDB aggregation pipeline, or Cypher statement. Otherwise, it remains at the deepest semantically valid position in the unifying layer. The implementation delegates the condition as a whole; it does not assume that partial translation of a compound condition preserves the same behavior.

This capability check is essential in a heterogeneous environment. The same conceptual filter may have different native representations, and a condition supported by one DBMS may not be supported by another. Predicate pushdown is therefore both mapping-aware and capability-aware rather than purely algebraic.

For an input containing \(N\) records and a filter with selectivity \(\rho\), where \(0 \leq \rho \leq 1\), successful pushdown may reduce the transferred result approximately from \(N\) to \(\rho N\) records. The reduction also propagates to subsequent joins and result-processing operators.

\begin{figure}[h]
    \centering
    \includegraphics[width=.48\textwidth]{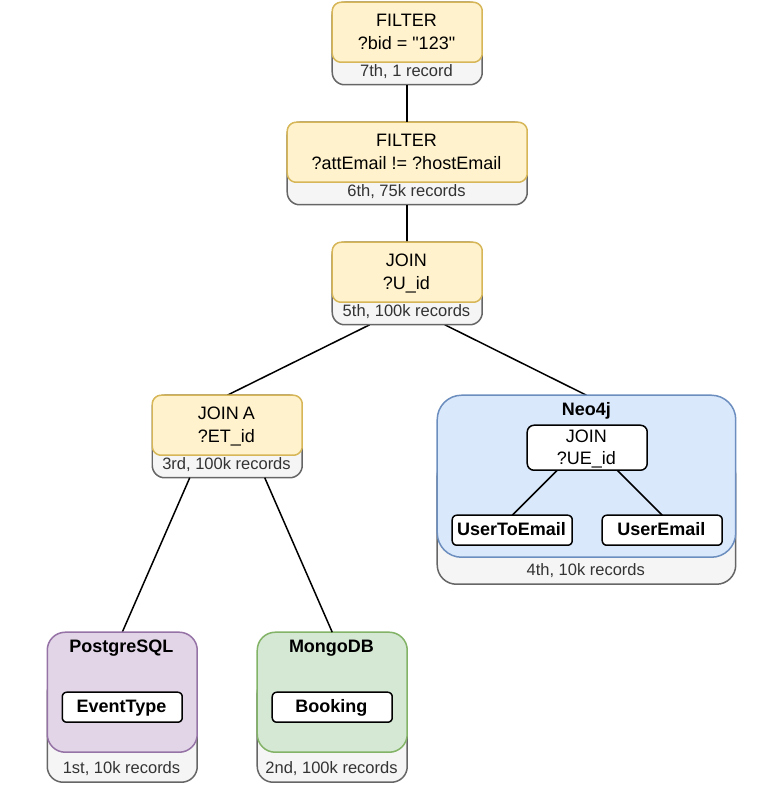}\hfill
    \includegraphics[width=.48\textwidth]{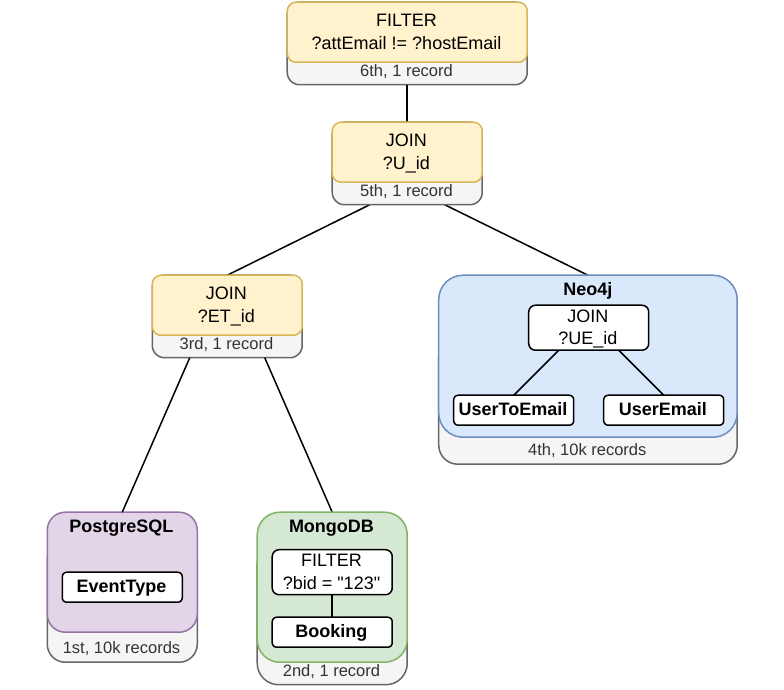}
    \caption{Predicate pushdown in the running example. Left: the original plan. Right: after pushing the applicable filter into the native query part.}
    \label{fig:predicate-pushdown}
\end{figure}

\begin{example}
Figure~\ref{fig:predicate-pushdown} compares the running-example QEP before and after predicate pushdown. The purple, green, and blue regions represent native query parts executed by PostgreSQL, MongoDB, and Neo4j, respectively; operations outside these regions are evaluated in the MM-quecat unifying layer. In the original plan on the left, the selective \texttt{?bid = "123"} filter remains outside the green MongoDB query part. MongoDB therefore returns approximately \(100\,000\) booking records, which are transferred and processed before the filter reduces them to one record. In the optimized plan on the right, the filter is incorporated into the green MongoDB query part, so only the matching booking is returned. This reduction propagates upward, substantially reducing the input to subsequent joins.
\qed
\end{example}

\subsection{Cross-Model Dependent Joins}
\label{sec:dependent-joins}

When a join cannot be delegated to one underlying DBMS, the standard execution strategy retrieves both inputs independently and evaluates the join in the unifying layer. Let \(A\) and \(B\) denote the two join inputs, and let \(N_A\) and \(N_B\) denote their actual cardinalities. This strategy retrieves and transfers both complete inputs, even when only a small subset of one input can contribute to the join result.

A dependent join introduces an execution dependency between the inputs. One input is evaluated first, and its join values are used to restrict the second input before it is retrieved. Variants of this idea are known in heterogeneous and distributed optimization under names such as bind joins, tuple-substitution joins, or related dependent execution schemes~\citep{opt-diverse-datasrc,opt-orthogonal}. Their common goal is to reduce remote work by using one branch to constrain another.

For an equality join on variable \(v\), the independent result produces the set
\[
D_v = \{v_1,\ldots,v_k\},
\]
where duplicate values are removed. The dependent input is then evaluated with an additional predicate equivalent to
\[
v \in D_v.
\]

After the restricted dependent result is retrieved, the original join is evaluated. For the supported inner equality joins, the additional predicate removes only records whose join value is absent from the independent input and therefore cannot change the final join result. Duplicate values are removed only from the generated restriction, not from either join input, so the multiplicities of the final result remain determined by the original join.

\subsubsection{Execution Model}

A dependent join introduces top-down data flow into an otherwise bottom-up query execution plan. The independent child must be evaluated before the dependent child, and values produced by the former must influence the native query generated for the latter.

We represent this dependency using an evaluation context. The dependent join first evaluates its independent child, extracts the distinct join values, and inserts the resulting restriction into the context. The dependent subtree is then evaluated with this context. Each visited node either applies the restriction, translates it into a native query part, or passes it further toward the leaves.

This representation allows the dependent input to be a complete subtree rather than only one data source leaf. The dependency is therefore defined at the model-independent plan level and may be realized by different native query languages at different leaves. A restriction may pass through intermediate plan operators until it reaches a query part that provides the join variable and supports the generated predicate.

\begin{figure}[h]
    \centering
    \includegraphics[width=.48\textwidth]{example-qep-2.pdf}\hfill
    \includegraphics[width=.48\textwidth]{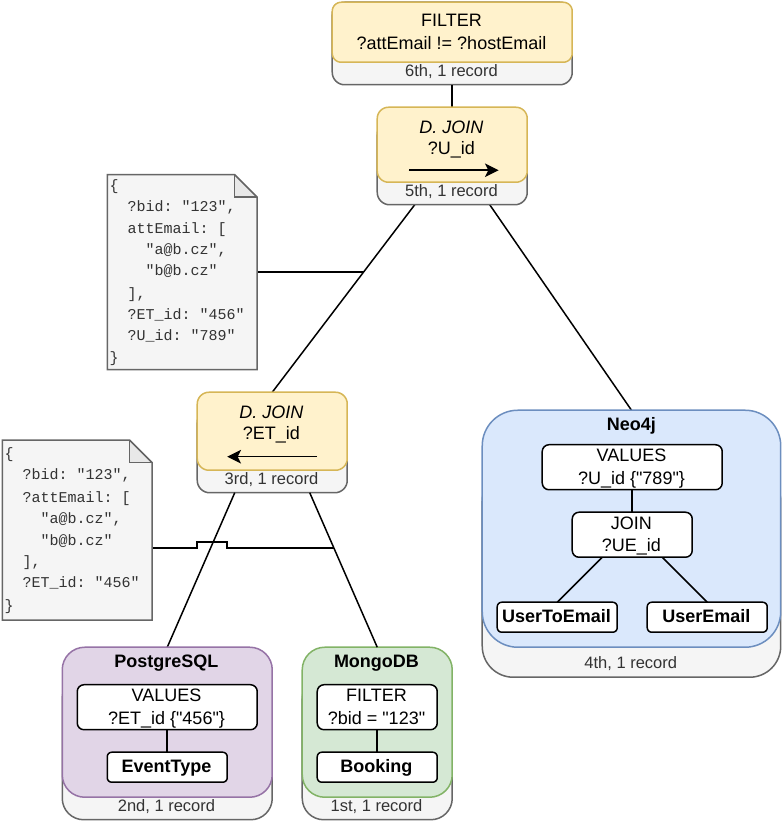}
    \caption{Dependent join in the running example. Left: the plan after predicate pushdown. Right: after introducing dependent joins that restrict the larger branches.}
    \label{fig:dependent-join}
\end{figure}

\begin{example}
Figure~\ref{fig:dependent-join} compares the plan after predicate pushdown on the left with the dependent-join plan on the right. The purple, green, and blue regions denote native PostgreSQL, MongoDB, and Neo4j query parts, respectively; operations outside these regions are evaluated in the MM-quecat unifying layer. The white boxes show intermediate records and the dynamically generated \texttt{VALUES} restrictions passed to dependent branches. In the right-hand plan, the joins on \texttt{?ET\_id} and \texttt{?U\_id} are replaced by dependent joins. The green MongoDB branch is evaluated first and produces the event-type value \texttt{"456"}, which is propagated as a restriction to the purple PostgreSQL branch before its result is retrieved. The resulting user identifier \texttt{"789"} is then propagated to the blue Neo4j branch. Consequently, both dependent branches return only records that can contribute to the final join result.
\qed
\end{example}

If the restriction cannot be delegated to the dependent subtree, the original non-dependent join is retained. The generated restriction is therefore an execution aid rather than a replacement for the join semantics, and the optimization preserves a safe fallback path.

\subsubsection{Orientation and Applicability}

The direction of a dependent join affects both the length of the generated predicate and its expected selectivity. Let \(\widehat{N}_A\) and \(\widehat{N}_B\) denote estimates of the actual cardinalities \(N_A\) and \(N_B\). When both orientations are possible, the input with the smaller estimated cardinality is selected as independent:
\[
\widehat{N}_A \leq \widehat{N}_B
\quad \Longrightarrow \quad
A \text{ is evaluated first}.
\]

This heuristic limits the number of values inserted into the dependent query and increases the probability that the larger input is reduced substantially.

Cardinality alone does not fully determine the benefit. The number of distinct join values, their distribution in the dependent source, and the cost of evaluating the additional predicate also matter. The current method, therefore, uses a conservative applicability condition. A dependent join is used only when:
\begin{enumerate}
    \item the join is an equality join;
    \item the dependent subtree can accept the generated filter;
    \item the estimated independent result does not exceed a configured limit;
    \item the resulting native query remains within the target DBMS's technical limits.
\end{enumerate}

The applicability condition is verified again during execution because a cardinality estimate may be inaccurate. If the actual independent result is too large, the processor falls back to the original join.

\subsubsection{Cardinality Information}

Dependent-join orientation requires estimates of intermediate-result cardinalities. Maintaining a complete statistical model for every relational, document, and graph database inside the unifying layer would duplicate functionality already provided by the underlying DBMSs and would require separate model-specific estimators.

The proposed approach, therefore, uses two sources of information. Native query builders may request available estimates from PostgreSQL, MongoDB, and Neo4j. Depending on the system, these may describe expected cardinality, native execution cost, or result size. Because the reported values have different meanings and scales, they are not treated as directly comparable global costs. This design choice is consistent with practical source-specific optimization, where local optimizers expose useful but heterogeneous planning signals rather than one shared global cost model~\citep{opt-diverse-datasrc,mongodb-optimizer}.

The second source consists of observations collected during previous executions. The processor records cardinalities and execution properties of query parts and reuses observations from structurally similar plans. The estimator is intentionally lightweight. Its purpose is to distinguish sufficiently clearly between smaller and larger join inputs, not to reproduce complete native cost models. This places it closer to query-driven estimation than to full statistical optimization and aligns it with broader work on adaptive and learned selectivity estimation~\citep{cost-query-driven-1,cost-query-driven-2}.

\subsection{Non-Redundant Query-Part Construction}
\label{sec:fast-query-parts}

Before a query execution plan can be created, MM-quecat identifies combinations of kind patterns whose properties cover the requested conceptual query. The original construction algorithm incrementally extends incomplete query parts. In every step, it calculates a redundancy score for the remaining candidate patterns and creates one continuation for each pattern with the minimum score.

This branching is necessary when candidates represent genuinely different physical alternatives. It is redundant, however, when several candidates are all mandatory. The general principle is the same as in classical optimizer design: equivalent alternatives should be explored only when they may lead to meaningfully different execution choices.

A kind pattern has a redundancy score \(1\) when it contains at least one required morphism that is not provided by any other remaining pattern. Such a pattern must occur in every complete extension of the current query part.

Let a construction state be represented by the pair \((s,Q)\), where \(s\) is the current incomplete query part and \(Q\) is the set of remaining candidate kind patterns. Let
\[
M = \{m_1,\ldots,m_k\} \subseteq Q
\]
be the set of patterns in \(Q\) with redundancy score \(1\). The original procedure creates one branch for every \(m_i\), after which it encounters the same situation with \(k-1\) patterns. It may therefore explore \(k!\) insertion orders, although every branch produces the same final set of kind patterns.

\begin{figure}[!t]
    \centering
    \includegraphics[width=0.9\textwidth]{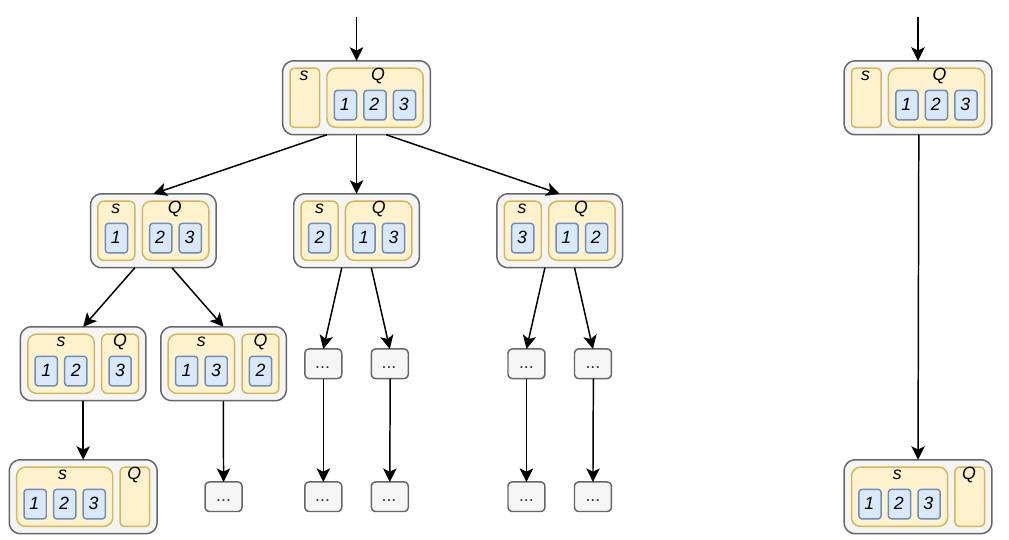}
    \caption{Original and non-redundant construction for three mandatory kind patterns. Each node represents a construction state \((s,Q)\), where \(s\) is the current incomplete query part and \(Q\) is the set of remaining candidate kind patterns. The labels \(1\), \(2\), and \(3\) denote the mandatory patterns. The original procedure explores their different insertion orders, whereas the optimized procedure transfers all three patterns from \(Q\) to \(s\) in a single step. The left-hand tree shows a prefix of the possible insertion-order paths.}
    \label{fig:query-part-construction}
\end{figure}

Figure~\ref{fig:query-part-construction} illustrates the difference between the original branching procedure and the joint addition of mandatory patterns.

The optimized procedure adds all patterns in \(M\) in one step, producing the successor state
\[
s' = s \cup M,
\qquad
Q' = Q \setminus M.
\]

Only one continuation is created. For \(k\) mandatory patterns, the original procedure may explore up to \(k!\) equivalent insertion orders, whereas the modified procedure creates one continuation containing all \(k\) patterns.

The method does not merge candidates with redundancy scores greater than \(1\), because such candidates may represent alternative mappings and are not necessarily required in every complete query part. Adding every score-one pattern is safe because each provides a required morphism unavailable from the other remaining candidates and must therefore occur in every complete extension of the current state. The optimization changes only the order in which these mandatory patterns are added; it removes provably redundant permutations while preserving all distinct physical alternatives.

\subsection{Optimization Scope}
\label{sec:optimization-scope}

The proposed techniques target inner select--project--join queries with filters over PostgreSQL, MongoDB, and Neo4j. Predicate pushdown and dependent joins are applied only when the required variables are covered by the selected mapping and the complete condition can be translated by the selected native query builder. Conditions for which the implementation cannot establish these properties remain in the unifying layer. Non-redundant query-part construction applies only to mandatory kind patterns identified by redundancy score \(1\).

Other possible reductions of the cross-model tax include earlier projection, aggregation, sorting, and limiting, more accurate global plan selection, and explicit modeling of operations unsupported by individual native query builders. These extensions require broader support for query languages and cost models and remain outside the scope of the evaluated implementation.
\section{Implementation in MM-quecat}
\label{sec:implementation}

The proposed techniques were implemented in MM-quecat~\citep{mmquecat_impl,mmquecat}, the Java-based query-processing component of MM-cat~\citep{mmcat_impl,mmcat}. The implementation extends the original bottom-up pipeline at four points: construction of physical query parts, representation and transformation of the query execution plan (QEP), collection of cardinality information, and generation of native queries. Table~\ref{tab:implementation-impact} summarizes these changes and separates planning-time, runtime, and native-query-generation responsibilities.

\begin{table}[h]
\centering
\scriptsize
\caption{Implementation impact of the proposed optimization techniques and the supporting cardinality estimator.}
\label{tab:implementation-impact}
\begin{tabular}{
>{\raggedright\arraybackslash}p{0.12\textwidth}
>{\raggedright\arraybackslash}p{0.18\textwidth}
>{\raggedright\arraybackslash}p{0.18\textwidth}
>{\raggedright\arraybackslash}p{0.20\textwidth}
>{\raggedright\arraybackslash}p{0.15\textwidth}
}
\toprule
\textbf{Technique} &
\textbf{QEP representation} &
\textbf{Planning} &
\textbf{Runtime execution} &
\textbf{Native-query generation} \\
\midrule
Non-redundant construction &
No QEP representation change &
Mandatory kind patterns are added in one expansion step &
None &
None \\
Predicate pushdown &
Filters can be stored either as explicit QEP nodes or inside datasource nodes &
Local downward transformation with variable and capability checks &
No new execution dependency &
Translation of static filter conditions \\
Cardinality estimation &
Execution observations are associated with structural plan properties &
Estimate lookup for candidate join inputs &
Collection of cardinality, execution-time, and transferred-volume observations &
Retrieval of source-specific planning information \\
Dependent joins &
Join nodes store the selected orientation and fallback behavior &
Orientation based on estimated input cardinalities &
Top-down restriction propagation through an evaluation context &
Translation of dynamically generated value restrictions \\
\bottomrule
\end{tabular}
\end{table}

The techniques are applied in the order in which they affect query processing. Non-redundant construction first limits the physical query parts considered by the planner. Predicate pushdown then rewrites the resulting QEP and moves translatable conditions into the datasource nodes. Cardinality information is used to orient eligible dependent joins, whose restrictions are generated and propagated during execution. All three techniques therefore share one QEP representation and the same translation boundary between model-independent processing and native SQL, MongoDB aggregation pipeline, or Cypher execution.

\subsection{Query-Plan Representation}
\label{sec:impl-qep}

In the original implementation, filters were represented exclusively as individual query execution plan (QEP) nodes. A native query part was represented by a datasource node whose subtree contained a pattern node and, potentially, filter nodes. This representation did not explicitly distinguish a filter retained in the unifying layer from one delegated to an underlying DBMS.

A datasource node was therefore extended to store both its pattern and the filters assigned to the native query part. A standalone filter node is evaluated in the unifying layer, whereas a filter stored inside a datasource node is translated and evaluated by the corresponding DBMS. This distinction makes operator placement explicit: predicate pushdown can change both the position of a filter in the QEP and the execution domain in which it is evaluated.

The QEP representation was also made mutable. Predicate pushdown consists of several local tree transformations, for which repeatedly constructing complete copies of the plan would introduce unnecessary planning overhead. The extended representation further provides the variables produced by each subtree, which are required when moving static filters and dynamically generated dependent restrictions.

These changes provide the common representation required by predicate pushdown, dependent-join planning, and native-query generation. They also make it possible to verify mapping coverage before a condition transitions from model-independent processing to a native query component.

\subsection{Native Query Generation}
\label{sec:impl-native-queries}

Native query builders translate query parts and model-independent conditions into SQL, MongoDB aggregation pipelines, and Cypher statements. Each builder first verifies that it supports the complete condition and only then constructs its native representation. The same translation interface is used for statically pushed filters and dynamically generated dependent restrictions; an unsupported condition remains outside the corresponding datasource node.

The PostgreSQL builder supports filtering and projection, and its construction of queries containing multiple joins was revised. A join is added only when it connects a kind already present in the partial query with a new kind. The available joins are processed repeatedly until all required kinds have been connected.

The MongoDB builder was extended to translate generic filter expressions and insert them as \texttt{\$match} stages in aggregation pipelines. This mechanism is used for both predicate pushdown and dependent restrictions.

The Neo4j builder translates mapped kinds, joins, filters, and projections into Cypher. Connected kinds and relationships are consolidated into graph patterns represented by as few \texttt{MATCH} clauses as possible, avoiding unnecessary fragmentation of one connected pattern. This builder-specific revision is implementation support and is not evaluated as a separate contribution.

The builders form the capability-aware translation boundary of the pipeline. Mapping coverage and builder support are separate requirements: an operation can be delegated only when the selected mapping provides all referenced values and the corresponding builder can translate the complete condition.

\subsection{Predicate Pushdown}
\label{sec:impl-pushdown}

Predicate pushdown is implemented by the \texttt{FilterDeepener} transformation, which is applied after the construction of the initial multi-model QEP. For each filter, the transformation repeatedly considers a deeper position. A move is accepted only when the target subtree provides every referenced variable and the intervening operator admits the transformation defined in Section~\ref{sec:predicate-pushdown}.

When a filter reaches a datasource node, the transformation checks whether the corresponding native builder supports the complete condition. If so, the standalone filter node is removed, and its condition is stored inside the datasource node. Otherwise, the filter remains at the deepest semantically valid position reached.

The transformation moves filters only toward the leaves and does not generate alternative QEPs. For a fixed input plan, it therefore produces one optimized plan rather than a new set of candidates. The evaluated implementation permits movement through inner joins; optional joins, set operations, and transformations depending on null preservation or unmatched-record semantics are left unchanged.

\subsection{Dependent Joins}
\label{sec:impl-dependent-joins}

Dependent joins introduce top-down information flow into the otherwise bottom-up QEP evaluator. Planning determines the dependency's orientation, while runtime execution obtains the concrete values used to restrict the dependent branch.

After predicate pushdown, the planner examines equality joins retained in the QEP and estimates their input cardinalities. A join is marked for dependent execution only when one input is estimated to remain below the configured limit, and the other subtree can accept a restriction over the join variable. When both orientations are eligible, the smaller estimated input is assigned to the independent branch.

At runtime, the independent branch is evaluated first. For a join variable \(v\), the implementation extracts the set of distinct values \(D_v\). The actual result size is checked because the planning estimate may be inaccurate. If it exceeds the permitted limit, the original non-dependent strategy is used.

Otherwise, a restriction over \(D_v\) is inserted into an evaluation context, and the dependent subtree is evaluated with this context. Intermediate QEP nodes propagate unresolved restrictions toward their descendants. A datasource node consumes a restriction only when its query part provides the join variable and its native builder can translate the generated condition.

This mechanism permits the dependent input to be a subtree rather than only a data source leaf. A restriction may pass through model-independent operators and be translated only when it reaches a suitable native query part. The original join is still evaluated after retrieval. If the restriction is too large or no datasource node can consume it, execution falls back to the original non-dependent strategy.

\subsection{Query-Driven Cardinality Estimation}
\label{sec:impl-cardinality}

The implementation uses a lightweight estimator designed for one specific decision: selecting the independent input of a dependent join. It does not attempt to implement a complete heterogeneous cost model.

During QEP execution, MM-quecat records execution time, result cardinality, and retrieved data volume for individual plan nodes. Observations are cached according to structural plan properties, including the accessed kinds and joins, while filters are handled separately. For a new plan, observations from structurally similar executions are retrieved and averaged. The estimator is used only to distinguish between smaller and larger candidate inputs for dependent-join orientation.

When no suitable execution observation is available, planning information can be requested from the underlying DBMSs through the Collector functionality. Collector provides a common interface to PostgreSQL, MongoDB, and Neo4j planning facilities. Its original interface combined several forms of query and database information; for MM-quecat, the integration was focused on extracting the source-specific information needed for estimation.

The systems expose various signals, including expected cardinalities, native execution costs, and result size indicators. Because these values have different semantics and scales, they are not directly comparable as a single global cost. They serve as source-specific fallback information when reusable execution observations are unavailable.

The estimator thus combines reusable execution observations with source-specific planning information while remaining deliberately task-specific. It neither normalizes the heterogeneous native estimates to a single cost scale nor selects the globally cheapest QEP.

\subsection{Non-Redundant Query-Part Construction}
\label{sec:impl-query-parts}

The query-part optimization was implemented in the expansion step of the construction algorithm. The original procedure identified all remaining candidate kind patterns with the minimum redundancy score and created a successor state for each.

The modified procedure first checks whether the minimum score is \(1\). In this case, every pattern with this score is mandatory because it provides a required schema fragment unavailable from any other remaining candidate. All such patterns are added to the current query part together and removed from the remaining candidate set before one successor state is created.

For minimum scores greater than \(1\), the original branching behavior is preserved because the candidates may represent genuinely different physical alternatives. The implementation, therefore, eliminates only insertion-order permutations of mandatory patterns.

General duplicate-state detection was not implemented. Different paths involving non-mandatory patterns may still reach equivalent intermediate states. The planning-time improvements reported in Section~\ref{sec:evaluation} therefore result specifically from the joint addition of mandatory patterns, not from complete deduplication of the search space.

\subsection{Implemented Scope}
\label{sec:impl-scope}

The evaluated implementation supports inner select--project--join queries with filters over PostgreSQL, MongoDB, and Neo4j. Predicate pushdown requires the target subtree to provide all referenced variables and the native builder to support the complete condition. Dependent joins are limited to equality conditions, translatable restrictions, and bounded independent results. Non-redundant construction applies only to mandatory kind patterns with redundancy score \(1\).

The implementation realizes the proposed mapping- and capability-aware pipeline across conceptual, mapping, and native execution levels. It is not a complete heterogeneous cost-based optimizer: unsupported transformations remain in the unifying layer, source-specific estimates are not treated as globally comparable costs, and ineligible dependent joins use the original execution strategy.
\section{Experimental Evaluation}
\label{sec:evaluation}

We evaluate whether the proposed techniques reduce the components of the cross-model tax identified in Section~\ref{sec:optimization}. The evaluation covers end-to-end latency, the main processing phases, the incremental effect of the optimization pipeline, and the query and mapping properties that contribute to its effectiveness. All conclusions refer to the evaluated MM-quecat implementation and deployment.

The experiments answer the following research questions:
\question{stages}{Which stages dominate the execution time of the original multi-model query-processing pipeline?}
\question{performance}{What is the incremental performance effect of predicate pushdown, dependent joins, and non-redundant query-part construction in the evaluated optimization pipeline?}
\question{properties}{Under which query and mapping properties are the individual techniques effective?}
\question{bottlenecks}{Which performance bottlenecks remain after all implemented optimizations are enabled?}

The evaluation uses MM-quecat as the experimental platform. The results characterize the proposed techniques within the evaluated decomposition-based architecture, dataset, supported MMQL subset, and single-machine containerized deployment. They are not intended as a general benchmark of multi-model database systems or as an evaluation of a physically distributed deployment.

\subsection{Experimental Setup}
\label{sec:eval-setup}

This subsection describes the experimental schema and mappings, the workload, and the execution protocol.

\subsubsection{Dataset and Mappings}

The current implementation primarily supports select--project--join queries with filters. We therefore base the experimental schema on Cal.com, an open-source scheduling application~\citep{caldotcom}, whose database schema is defined using the Prisma object-relational mapping framework~\citep{prisma}.

We converted a connected subset of the Cal.com schema into an MM-cat schema category containing 70 objects and 90 morphisms. The subset emphasizes relationships between complex objects rather than scalar attributes, and its maximum schema graph degree is 6. It consequently produces queries that differ both in conceptual size and in the number of physical kinds required under different mappings.

Complete mappings were created for PostgreSQL, MongoDB, and Neo4j. The MongoDB representation uses 8 mappings, each containing nested and array-valued properties. The PostgreSQL representation uses 25 mappings, whereas the Neo4j representation uses 55 because graph relationships are represented as distinct types. Identifier properties used for references and joins were indexed; the remaining properties were not indexed.

We constructed four execution environments:

\begin{enumerate}
    \item a PostgreSQL-only environment;
    \item a MongoDB-only environment;
    \item a Neo4j-only environment;
    \item a heterogeneous environment in which the three DBMSs cover disjoint parts of the schema.
\end{enumerate}

The three single-DBMS environments represent the complete selected schema within a single physical model. The heterogeneous environment deliberately contains no redundant mappings. It therefore isolates cross-system execution without introducing additional mapping alternatives caused by replicated representations.

The data were generated synthetically and loaded into all physical representations. A dedicated initialization tool first generates conceptual records in memory, creates the physical kinds required by each mapping, transforms the relevant record subsets into the target representation, and loads them into the corresponding DBMS. The generator uses non-uniform value distributions over shuffled records to approximate selected properties of transactional workloads. It does not reproduce all correlations, skew, or irregularities of production Cal.com data. The generated dataset contains approximately 2,000,000 conceptual records.

% The generated dataset contains approximately \emph{2,000,000} conceptual records. After loading, the PostgreSQL representation contains \emph{1,935,867} rows with its largest table containing \emph{300,000} rows, the MongoDB representation contains \emph{451,321} documents with its largest collection containing \emph{240,000} rows, and the Neo4j representation contains \emph{4,959,533} nodes and relationships, with its most numerous node or relationship tag containing \emph{300,000} instances. The physical database sizes of all representations are approximately \emph{250 MB}.

\subsubsection{Workload}

The workload contains 20 MMQL queries derived from query structures found in the Cal.com codebase. They are predominantly parameterized select--project--join queries with equality, inequality, or set-membership predicates and joins along reference morphisms.

A language model was used to generate an initial set of candidate queries, following the workload-generation approach outlined in SQLStorm~\citep{sqlstorm}. The candidates were subsequently inspected: unsupported, redundant, and unsuitable queries were removed, and the workload was completed manually with more complex queries and queries containing larger graph patterns. The final workload is therefore neither a production trace nor a directly generated benchmark. It is a controlled workload combining application-derived structures with manual additions selected to exercise the supported optimization cases.

Table~\ref{tab:queries} summarizes the workload. The \emph{Filter restrictiveness} column is a qualitative classification rather than a measured selectivity: a higher value means that a smaller fraction of the input is expected to remain after filtering. The last four columns show the number of physical kinds accessed in the PostgreSQL, MongoDB, Neo4j, and heterogeneous environments. These values demonstrate that the same conceptual query may induce substantially different physical decompositions.

\begin{table}[h]
    \centering
    \scriptsize
    \begin{tabular}{ccccccc}
        \toprule
        \textbf{Query} & \textbf{Objects} & \textbf{Filter restrictiveness} & \textbf{$K_P$} & \textbf{$K_M$} & \textbf{$K_N$} & \textbf{$K_H$} \\
        \midrule
        Q0  & 6  & high   & 1 & 1 & 2 & 1 \\
        Q1  & 5  & low    & 2 & 1 & 3 & 1 \\
        Q2  & 4  & high   & 1 & 1 & 2 & 2 \\
        Q3  & 7  & medium & 3 & 1 & 5 & 5 \\
        Q4  & 5  & medium & 2 & 1 & 3 & 3 \\
        Q5  & 8  & high   & 2 & 1 & 4 & 4 \\
        Q6  & 7  & low    & 2 & 1 & 4 & 1 \\
        Q7  & 7  & high   & 3 & 1 & 5 & 5 \\
        Q8  & 7  & low    & 3 & 1 & 5 & 3 \\
        Q9  & 8  & medium & 2 & 1 & 3 & 2 \\
        Q10 & 6  & high   & 1 & 2 & 3 & 1 \\
        Q11 & 7  & low    & 3 & 2 & 5 & 5 \\
        Q12 & 9  & low    & 4 & 3 & 7 & 4 \\
        Q13 & 6  & medium & 3 & 2 & 5 & 2 \\
        Q14 & 8  & medium & 3 & 2 & 6 & 6 \\
        Q15 & 10 & low    & 4 & 3 & 6 & -- \\
        Q16 & 7  & low    & 4 & 3 & 5 & 4 \\
        Q17 & 7  & low    & 4 & 3 & 6 & -- \\
        Q18 & 9  & low    & 5 & 3 & 8 & -- \\
        Q19 & 8  & low    & 4 & 3 & 7 & 4 \\
        \bottomrule
    \end{tabular}
    \caption{Properties of the experimental workload. $K_P$, $K_M$, $K_N$, and $K_H$ denote the numbers of accessed kinds in the PostgreSQL, MongoDB, Neo4j, and heterogeneous environments, respectively.}
    \label{tab:queries}
\end{table}

Queries Q15, Q17, and Q18 could not be executed successfully in the heterogeneous environment due to functional errors in the prototype's execution path. These failures occur already in the original configuration and are not introduced by the evaluated optimizations. The three affected query--environment combinations are excluded from the corresponding comparisons rather than replaced by estimated values. All other combinations completed without functional errors, although several unoptimized executions exhausted the available memory, as discussed in Section~\ref{sec:eval-baseline}.

\subsubsection{Execution Protocol}

The performance tests were executed on the MM-cat application server rather than via Maven, so that the measurements reflect the framework's normal deployment path.

For each query, we measure total latency and the time spent in four processing phases:

\begin{enumerate}
    \item query planning;
    \item native query execution and transfer from the underlying DBMSs;
    \item joining and processing of intermediate results;
    \item final construction of the returned result.
\end{enumerate}

Native execution and transfer are measured together because both are handled through the database drivers and may overlap. The measurements, therefore, separate work performed in the unifying layer from the combined cost of native processing and result retrieval, but they do not isolate network transfer from DBMS execution.

Each query is executed four times for warm-up and twenty additional times for measurement. Query parameters are selected from stored values so that parameterized predicates match existing records. The DBMS containers and the Java process remain warm between measurements. Preliminary runs showed that latencies stabilized after the fourth execution, which determined the warm-up length. The figures report arithmetic means over the twenty measured executions. Because they do not show full latency distributions or confidence intervals, the values are interpreted as measurements of the evaluated deployment rather than as population estimates.

\begin{table}[h]
    \centering
    \scriptsize
    \begin{tabular}{ll}
        \toprule
        \multicolumn{2}{c}{\textbf{Hardware}} \\
        \midrule
        CPU & AMD Ryzen 7 4800H \\
        Memory & 32 GB \\
        \midrule
        \multicolumn{2}{c}{\textbf{Software}} \\
        \midrule
        Operating system & Arch Linux 2026.07.01 \\
        Java & OpenJDK 21.0.11 \\
        Docker & 29.6.1 \\
        PostgreSQL & 18.0 \\
        MongoDB & 8.2.1 \\
        Neo4j & 2025.10.1 \\
        \midrule
        \multicolumn{2}{c}{\textbf{Protocol}} \\
        \midrule
        Warm-up executions & 4 \\
        Measured executions & 20 \\
        Erroneous executions & excluded \\
        DBMS state & kept warm \\
        \bottomrule
    \end{tabular}
    \caption{Experimental environment and execution protocol.}
    \label{tab:environment}
\end{table}

The experiments use four cumulative configurations:

\begin{enumerate}
    \item the original implementation;
    \item predicate pushdown enabled;
    \item predicate pushdown and dependent joins enabled;
    \item all implemented optimizations enabled.
\end{enumerate}

The difference between two successive configurations represents the incremental effect of the newly enabled technique, given the preceding optimizations. The evaluation, therefore, measures the complete pipeline in its intended order, but it is not a factorial study of all possible optimization combinations. In particular, the reported effect of dependent joins assumes that predicate pushdown is already enabled.

\subsection{Baseline Performance and Bottlenecks}
\label{sec:eval-baseline}

\begin{figure}[h]
    \centering
    \includegraphics[width=\textwidth]{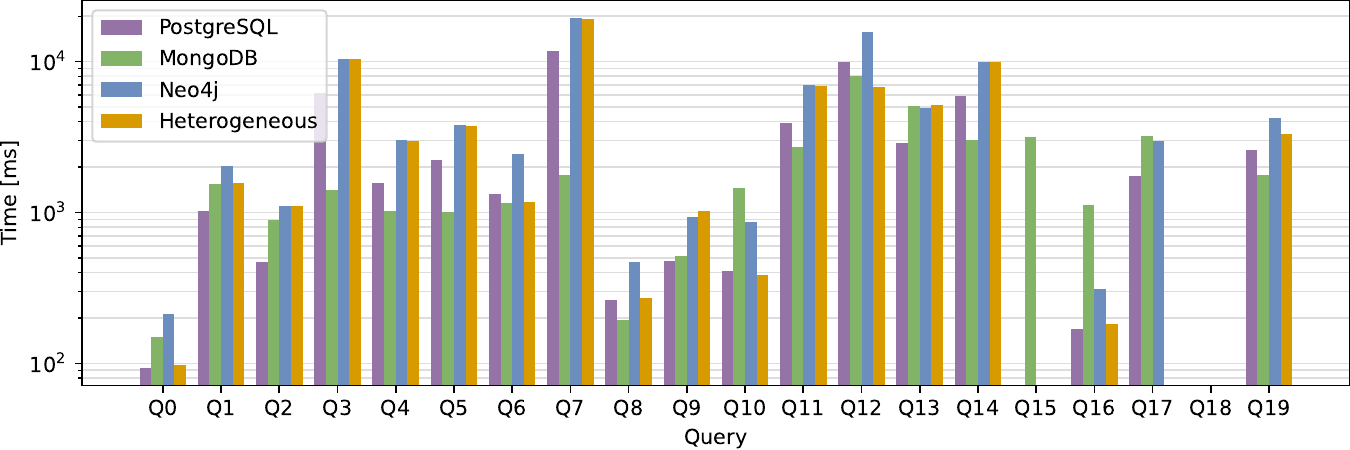}
    \caption{Mean query latency in the four environments before optimization. Missing values correspond to failed executions.}
    \label{fig:baseline-all}
\end{figure}

Before optimization, mean query latency ranged from less than ten milliseconds to more than ten seconds. Figure~\ref{fig:baseline-all} compares the four environments.

The large differences between queries cannot be explained by conceptual query size alone. They also reflect physical decomposition, the amount of data returned by native query parts, and the amount of joining retained in MM-quecat. PostgreSQL and MongoDB provide the lowest latency for different queries, while Neo4j is generally slower in the original implementation. The heterogeneous environment is not systematically faster than single-DBMS environments; when their latencies are similar, the corresponding query is often executed by the same underlying DBMS in both environments.

Some executions of Q15 and Q18 exhaust the memory available to MM-quecat. In the heterogeneous environment, Q15, Q17, and Q18 fail already in the original prototype execution path and therefore cannot be included in the optimization comparisons.

\begin{table}[h]
    \centering
    \scriptsize
    \begin{tabular}{ccccc}
        \toprule
        \textbf{Query} & \textbf{PostgreSQL} & \textbf{MongoDB} & \textbf{Neo4j} & \textbf{Heterogeneous} \\
        \midrule
        Q15 & OOM & OK & OOM & error \\
        Q17 & OK & OK & OK & error \\
        Q18 & OOM & OOM & OOM & error \\
        \bottomrule
    \end{tabular}
    \caption{Failed executions in the original configuration.}
    \label{tab:failed}
\end{table}

\begin{figure}[h]
    \centering
    \includegraphics[width=\textwidth]{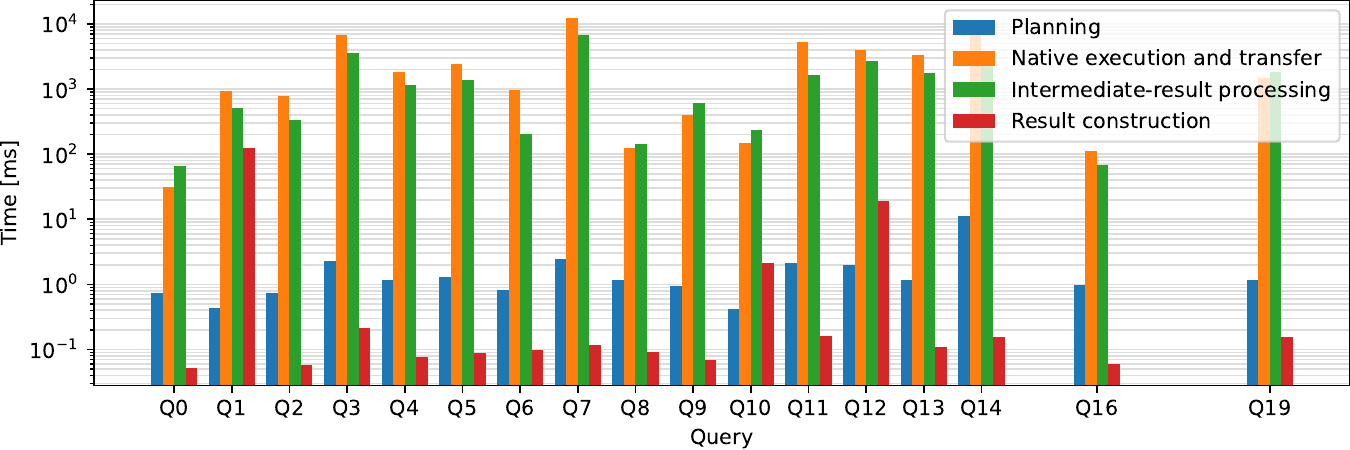}
    \caption{Mean time spent in individual query-processing phases in the unoptimized heterogeneous environment.}
    \label{fig:baseline-phases}
\end{figure}

The phase-level measurements identify two dominant components: native query execution and transfer, and processing of intermediate results in MM-quecat. Planning and final result construction are generally at least one order of magnitude smaller. Figure~\ref{fig:baseline-phases} illustrates this behavior in the heterogeneous environment.

These results answer \questionRef{stages}. In the evaluated deployment, the original pipeline is primarily limited by the volume of data retrieved from the underlying systems and subsequently converted and processed within the unifying layer. Predicate pushdown and dependent joins directly target these dominant components of the cross-model tax. Query-part construction becomes relevant for plans decomposed into larger numbers of physical kinds.

\subsection{Overall Optimization Impact}
\label{sec:eval-overall}

\begin{figure*}
    \centering
    \includegraphics[width=.82\textwidth]{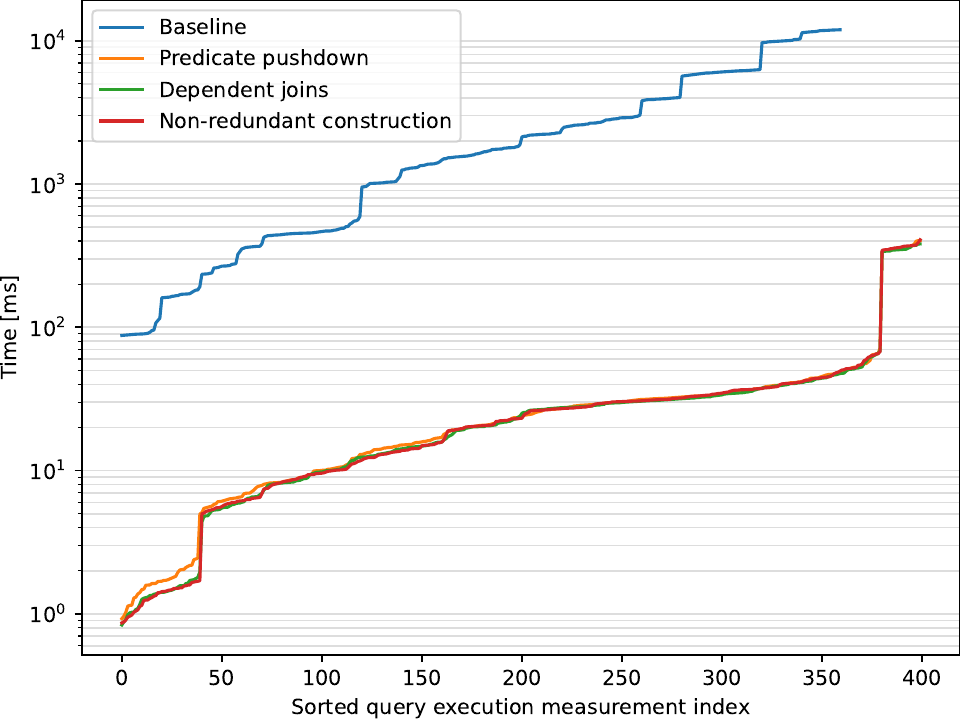}
    \caption{Cumulative optimization impact in the PostgreSQL environment.}
    \label{fig:impact-postgresql}
\end{figure*}

\begin{figure*}
    \centering
    \includegraphics[width=.82\textwidth]{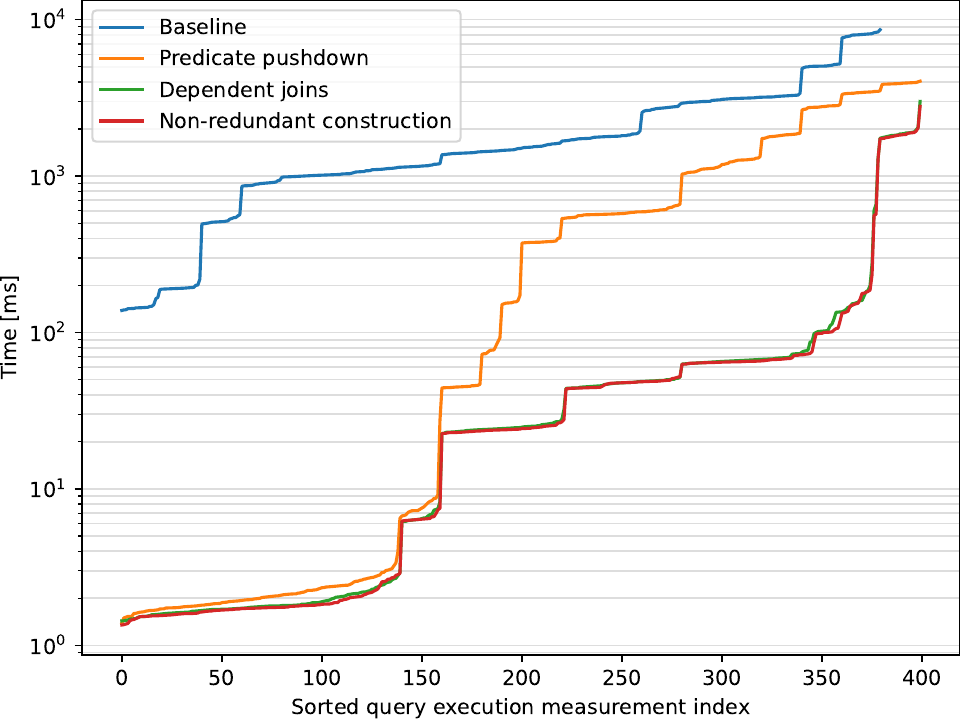}
    \caption{Cumulative optimization impact in the MongoDB environment.}
    \label{fig:impact-mongodb}
\end{figure*}

\begin{figure*}
    \centering
    \includegraphics[width=.82\textwidth]{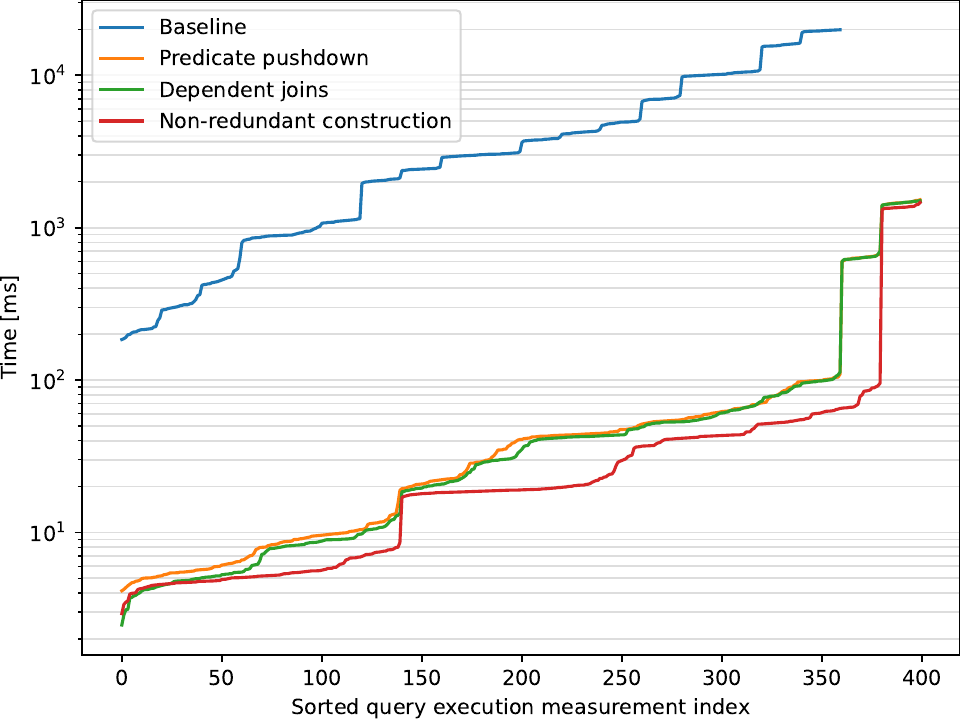}
    \caption{Cumulative optimization impact in the Neo4j environment.}
    \label{fig:impact-neo4j}
\end{figure*}

\begin{figure*}
    \centering
    \includegraphics[width=.82\textwidth]{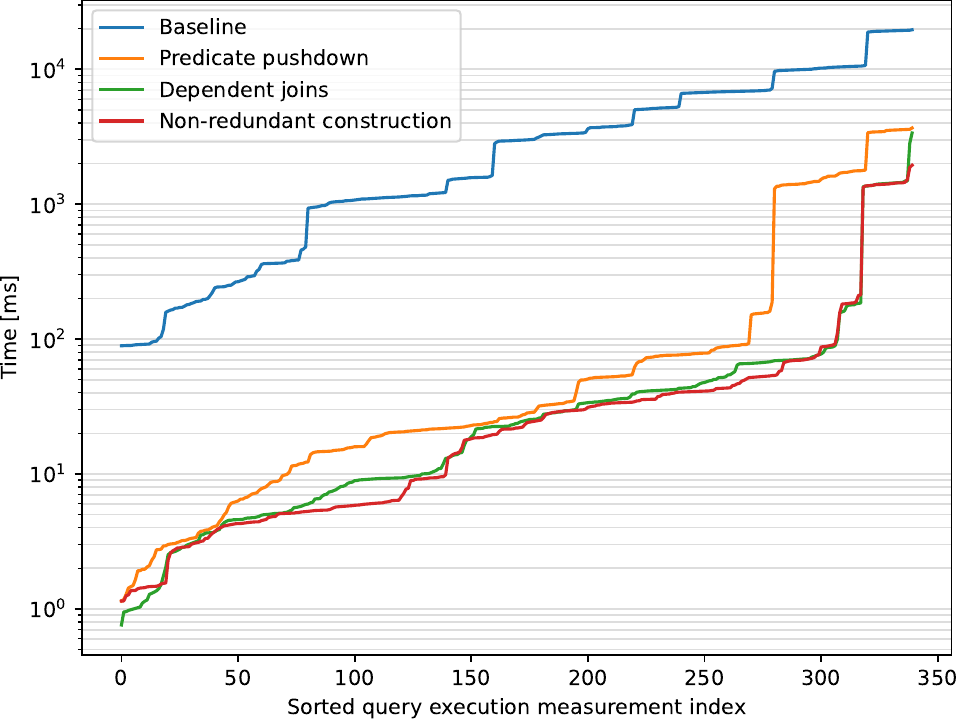}
    \caption{Cumulative optimization impact in the heterogeneous environment.}
    \label{fig:impact-heterogeneous}
\end{figure*}

Figures~\ref{fig:impact-postgresql}--\ref{fig:impact-heterogeneous} compare latency distributions under the four cumulative configurations. Each curve is sorted independently; the figures therefore compare distributions rather than matching queries or executions point by point.

The comparison also shows that the effectiveness of optimization depends on the physical decomposition induced by the mappings. Predicate pushdown has a broad effect across all environments, whereas dependent joins and non-redundant construction affect only plans in which joins or larger groups of mandatory kinds remain outside a single native query.

\subsubsection{Predicate Pushdown}

Predicate pushdown provides the broadest improvement because every workload query contains at least one filter. For the query--environment combinations most affected by large intermediate results, the maximum observed latency reduction reaches two orders of magnitude; this maximum is not representative of every workload query.

The largest improvements occur when a restrictive predicate can be incorporated into a native query, and the remaining operations can also be executed by the same DBMS. Pushdown then reduces not only native output, but also transfer, conversion, joining, and subsequent result processing.

Predicate pushdown also prevents all out-of-memory failures observed in the original single-DBMS experiments. These failures were caused by excessive intermediate results rather than by native DBMS execution itself. Reducing the results before they enter MM-quecat, therefore, changes several queries from unsuccessful to executable, improving robustness in addition to latency.

The improvement is smaller for several complex MongoDB and heterogeneous queries. Their filters reduce local query parts, but joins remain in the unifying layer and their inputs remain relatively large. Predicate pushdown can therefore reduce a cross-model join without necessarily eliminating its dominant cost.

\subsubsection{Dependent Joins}

Dependent joins do not affect the evaluated PostgreSQL and Neo4j plans because their relevant joins are already incorporated into native queries. They improve selected MongoDB and heterogeneous queries whose decompositions retain joins in MM-quecat.

For eligible query--environment combinations, the maximum observed additional improvement is approximately one order of magnitude. The technique is most effective when one of the input joins is small, and its distinct values strongly restrict the other input. Its benefit is therefore determined by the physical decomposition and data distribution, not merely by the presence of a conceptual join.

Dependent execution is not applicable to every external join. For Q12, both inputs contain several thousand records. The generated restriction would exceed the configured or DBMS-imposed limit, so MM-quecat retains the ordinary join. The conservative applicability condition prevents an expensive intermediate result from being replaced by an excessively large native predicate.

\subsubsection{Non-Redundant Query-Part Construction}

Non-redundant construction has little visible effect in the PostgreSQL, MongoDB, and heterogeneous environments, where the evaluated plans access at most 5, 3, and 6 kinds, respectively.

Its effect is strongest in Neo4j, whose mappings produce plans containing up to eight kinds. For the largest evaluated plans, total latency improves by approximately one order of magnitude, primarily due to reduced planning time. Section~\ref{sec:eval-planning} analyzes this effect separately.

These results answer \questionRef{performance}. Predicate pushdown reduces execution overhead across the broadest set of evaluated queries and prevents failures caused by excessive intermediate results. With pushdown already enabled, dependent joins provide further improvements for selected externally evaluated joins. Non-redundant construction reduces planning overhead when a conceptual query is decomposed into a larger number of mandatory physical kinds. Because the configurations are cumulative, these statements describe incremental effects in the order of evaluation rather than the independent effects of all possible combinations.

\subsection{Per-Query Behavior}
\label{sec:eval-per-query}

\begin{figure}[h]
    \centering
    \includegraphics[width=\textwidth]{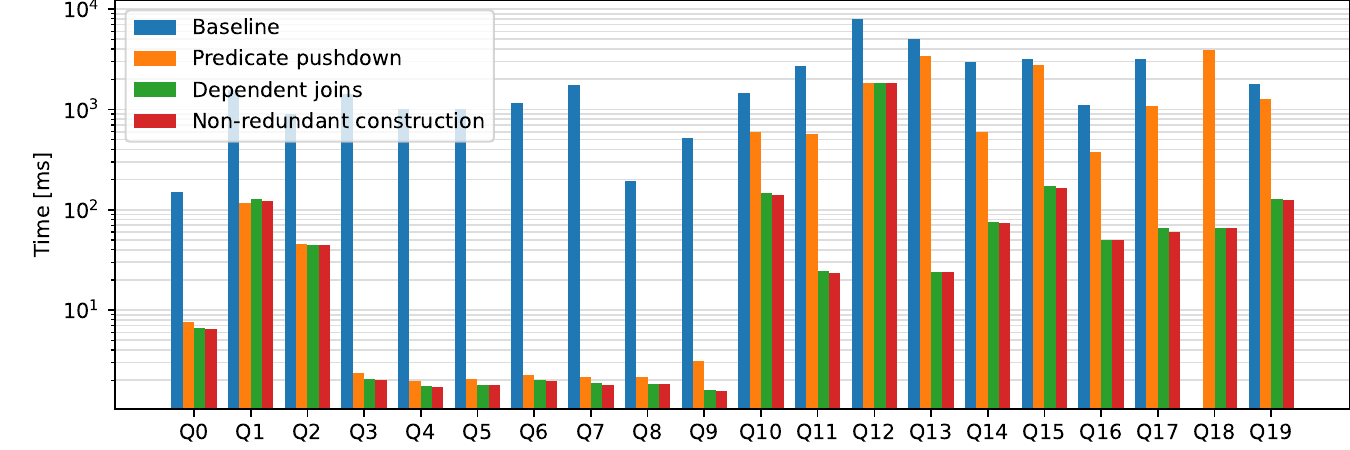}
    \caption{Per-query effects of the cumulative optimizations in the MongoDB environment.}
    \label{fig:per-query-mongodb}
\end{figure}

\begin{figure}[h]
    \centering
    \includegraphics[width=\textwidth]{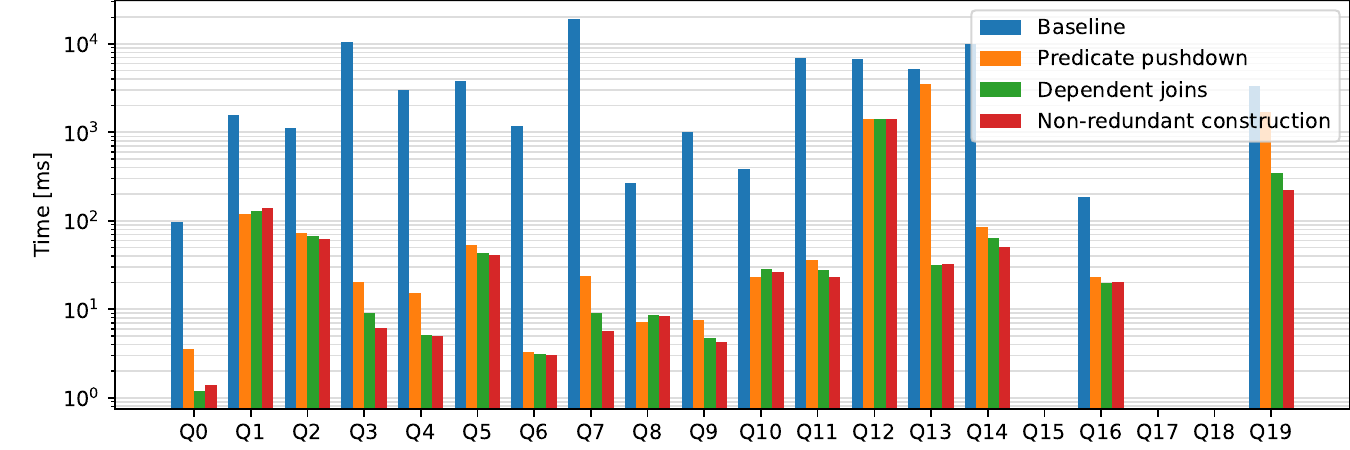}
    \caption{Per-query effects of the cumulative optimizations in the heterogeneous environment.}
    \label{fig:per-query-heterogeneous}
\end{figure}

\begin{figure}[h]
    \centering
    \includegraphics[width=\textwidth]{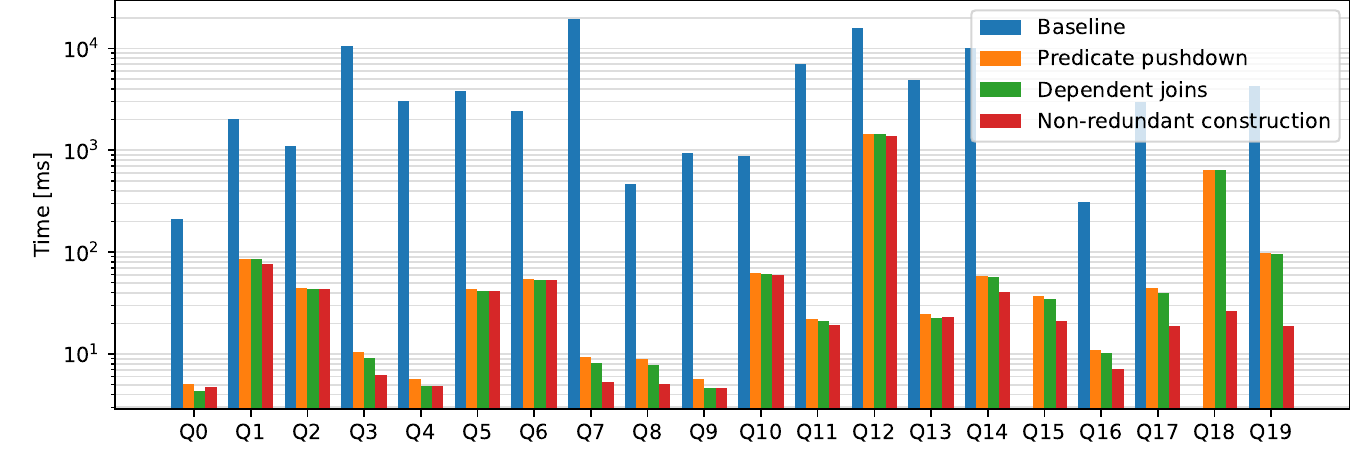}
    \caption{Per-query effects of the cumulative optimizations in the Neo4j environment.}
    \label{fig:per-query-neo4j}
\end{figure}

The cumulative distributions show the overall effect but do not identify which queries are affected. Figures~\ref{fig:per-query-mongodb}--\ref{fig:per-query-neo4j} therefore compare the configurations per query.

Figure~\ref{fig:per-query-mongodb} shows the MongoDB environment. Dependent joins primarily improve Q10--Q19 because these queries require joins across multiple MongoDB kinds. Q12 is the main exception because its independent result is too large for the generated restriction.

Figure~\ref{fig:per-query-heterogeneous} shows the heterogeneous environment. Different queries benefit from heterogeneous mappings, which induce different decompositions and cross-system joins. The applicability of optimization, therefore, cannot be inferred from the conceptual query alone.

Figure~\ref{fig:per-query-neo4j} shows the Neo4j environment. The main additional benefit comes from the non-redundant construction of query parts. Q18 and Q19 show the strongest reductions because they access 8 and 7 physical graph kinds, respectively.

These observations answer \questionRef{properties}. The effect of each technique is determined by a combination of conceptual query properties, mapping structure, and native capabilities:

\begin{itemize}
    \item predicate pushdown benefits queries with restrictive filters that can be translated by the selected native query builder;
    \item dependent joins benefit externally evaluated equality joins with one sufficiently small input and a restrictive set of join values;
    \item non-redundant construction benefits queries decomposed into larger numbers of mandatory physical kinds.
\end{itemize}

Optimization applicability is therefore a property of the conceptual query together with its physical realization, not of the conceptual query in isolation. The same MMQL query may require different techniques under different mappings.

\subsection{Planning-Time Impact}
\label{sec:eval-planning}

\begin{figure}[h]
    \centering
    \includegraphics[width=0.82\textwidth]{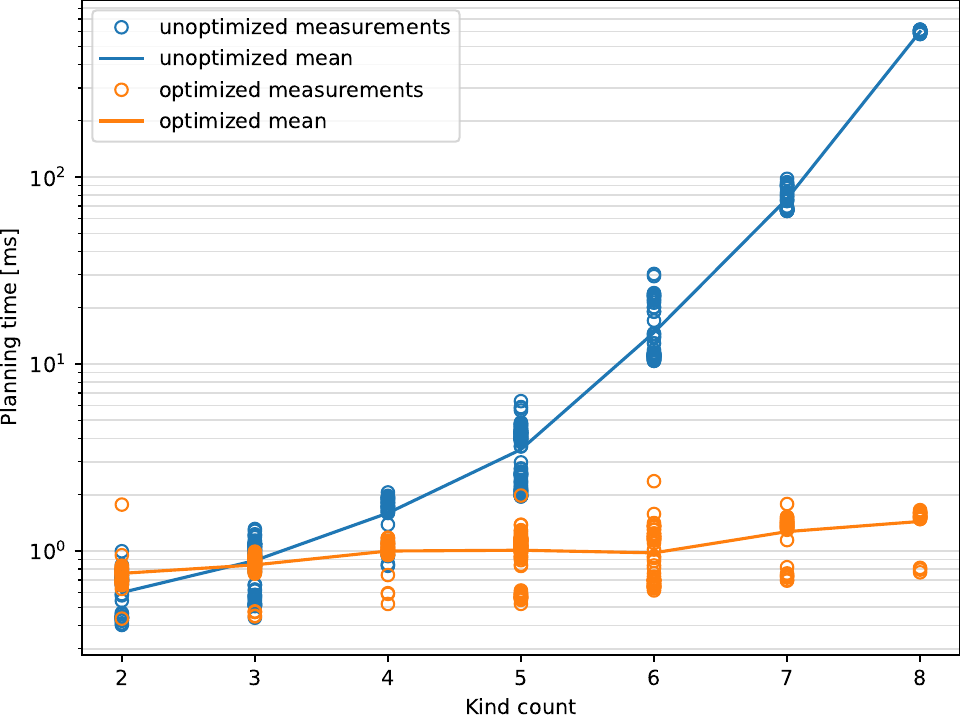}
    \caption{Planning time with respect to the number of queried Neo4j kinds before and after non-redundant query-part construction.}
    \label{fig:planning-time}
\end{figure}

Figure~\ref{fig:planning-time} relates planning time to the number of Neo4j kinds accessed by the evaluated queries. Planning time increases substantially for plans involving more kinds, reaching almost 600 milliseconds for the evaluated eight-kind plan. After mandatory patterns are added jointly, even the largest evaluated plans are constructed in several milliseconds.

The result is consistent with the combinatorial behavior analyzed in Section~\ref{sec:fast-query-parts}. Planning is not a major component of smaller plans, where saving several milliseconds may be overshadowed by longer execution times. It becomes significant when one representation decomposes a conceptual query into many mandatory kinds.

The optimized procedure removes the redundant insertion-order permutations in the evaluated mandatory-pattern cases. Because the queries differ in more properties than their number of accessed kinds, Figure~\ref{fig:planning-time} is a comparison of the evaluated plans rather than a controlled scalability law.

\subsection{Performance after Optimization}
\label{sec:eval-final}

\begin{figure}[h]
    \centering
    \includegraphics[width=\textwidth]{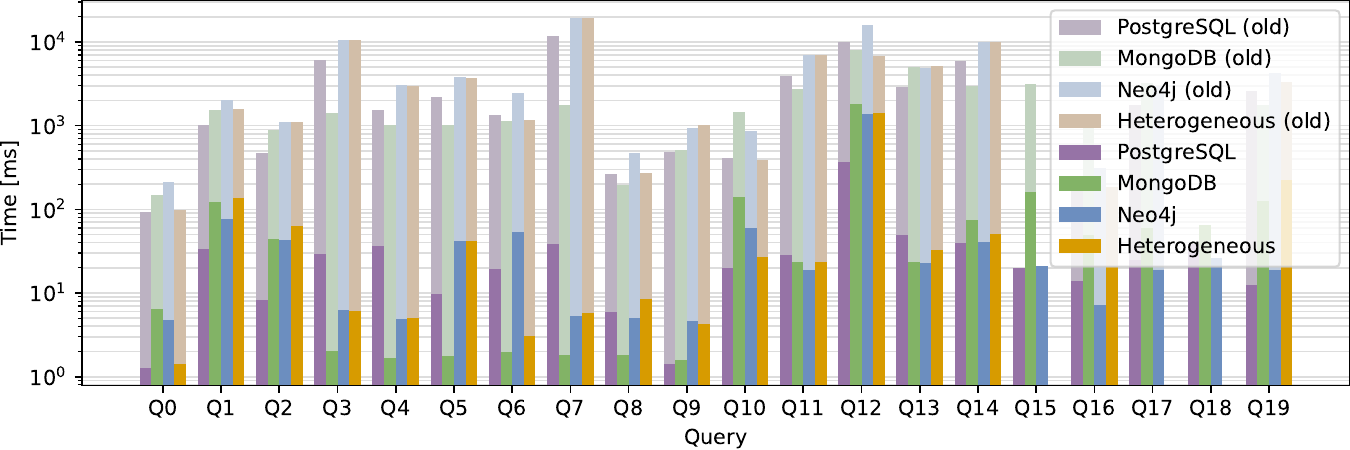}
    \caption{Mean query latency in the original and fully optimized configurations. Series marked as ``old'' represent the original configuration; the remaining series represent the fully optimized configuration.}
    \label{fig:optimized-all}
\end{figure}

Figure~\ref{fig:optimized-all} compares the original and fully optimized configurations. Most successful optimized executions require from several milliseconds to several hundred milliseconds, while Q12 remains above one second in some environments.

For some queries, predicate pushdown allows the entire operation to be delegated to a single DBMS, eliminating intermediate-result processing in MM-quecat. For more complex queries, native execution and transfer remain dominant. This is the intended effect of the pipeline: work is moved from the generic unifying layer to specialized engines closer to the data.

Unlike in the original configuration, each single-DBMS environment provides the lowest latency for at least some queries. The results, therefore, also show that relational, document, and graph representations can each be advantageous for particular physical query structures once the unifying layer no longer obscures their native strengths through excessive intermediate processing.

The main remaining expensive cases require several native query parts and externally evaluated joins. Q12 remains expensive because neither of the join inputs is sufficiently small to support a dependent restriction.

These results answer \questionRef{bottlenecks}. After optimization, the main remaining bottleneck is cross-part joining when neither input can efficiently restrict the other. Further improvements require broader join ordering, source and mapping selection, or alternative strategies for transferring and combining large intermediate results.

\subsection{Implications for Mapping Selection}
\label{sec:eval-oracle}

\begin{figure}[h]
    \centering
    \includegraphics[width=0.82\textwidth]{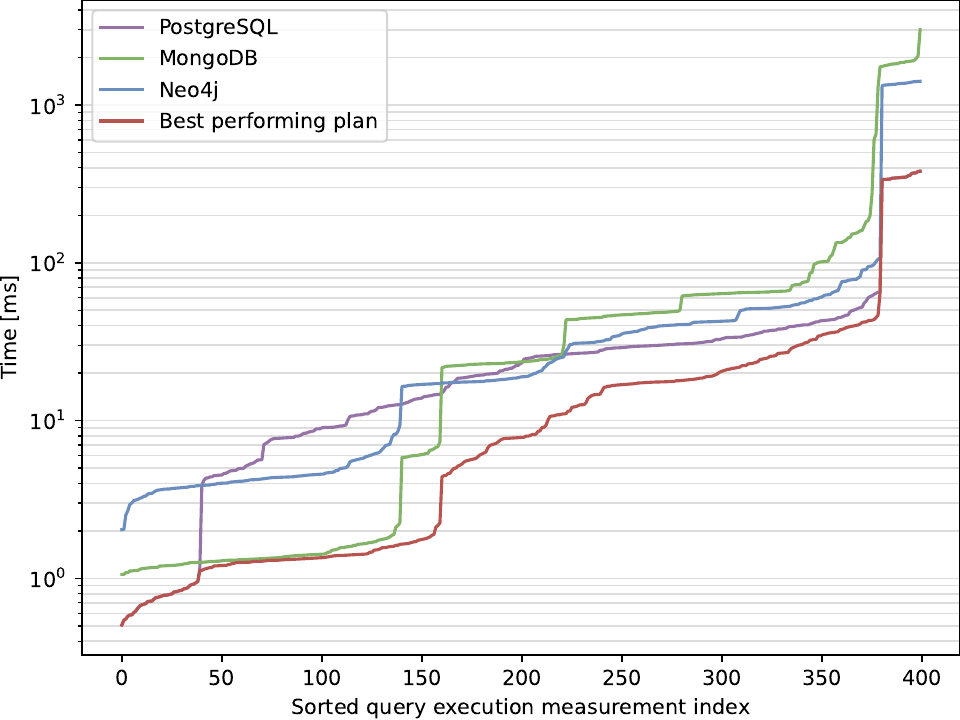}
    \caption{Optimized single-DBMS performance and a post-hoc oracle selecting the environment with the lowest mean latency for each query.}
    \label{fig:oracle}
\end{figure}

The three single-DBMS environments additionally illustrate the potential benefit of selecting among alternative physical representations. Figure~\ref{fig:oracle} reports a post-hoc comparison derived from the measured means; it is not an implemented runtime-selection strategy and is not included among the contributions evaluated in this paper.

For each query, the oracle selects the single-DBMS environment with the lowest measured mean latency. It reduces workload-wide mean latency by approximately \(20\%\) relative to the best fixed single-DBMS environment. This result is an upper bound derived from observed executions rather than achieved runtime-selection performance.

Within the evaluated workload, no single physical representation is uniformly best. This observation motivates cost-based selection among redundant mappings as a possible next step complementary to the operator-placement and plan-construction techniques evaluated here; it does not demonstrate the performance of such a selection method.

Taken together, the results support one consistent design principle: a decomposition-based multi-model processor should minimize work performed after native query boundaries. The most effective techniques either reduce data before it crosses these boundaries, use one branch to restrict another before retrieval, or avoid constructing redundant alternatives before execution begins.

\subsection{Threats to Validity}
\label{sec:threats}

The evaluation has several limitations.

\subsubsection{Dataset and workload}

The experimental data are generated from a real application schema but do not reproduce the full distributions, correlations, and irregularities observed in production Cal.com data. The workload contains 20 read-only queries and covers the MMQL subset currently supported by MM-quecat. Updates, complex aggregations, optional joins, and analytical workloads are not evaluated.

The initial workload generation involved a language model, although unsupported and redundant queries were removed and the final workload was manually completed. It should therefore be interpreted as a controlled workload inspired by application query structures rather than a trace of production executions.

\subsubsection{Experimental environment}

All systems run in Docker containers on one physical machine. The measurements include interprocess communication, driver overhead, conversion, and serialization, but not the higher and more variable latency of a physically distributed deployment. Such a deployment could increase the importance of reducing the transfer of intermediate results.

Conversely, the single-machine setting may introduce resource contention between MM-quecat and the participating DBMSs. The measured values characterize the complete selected deployment rather than isolated DBMS performance.

\subsubsection{Measurement and optimization isolation}

The reported values are means over 20 measured executions, following 4 warm-up executions. The figures do not report full latency distributions or confidence intervals, so the quantitative results characterize the measured runs and should not be interpreted as population-level performance estimates.

The optimization configurations are cumulative. Differences between successive configurations show incremental effects in one fixed order, but the evaluation is not a complete factorial analysis. In particular, the reported effect of dependent joins assumes that predicate pushdown is already enabled.

\subsubsection{Prototype limitations}

Three of the twenty workload queries fail in the heterogeneous environment because of implementation errors. Several original single-DBMS executions fail because of memory exhaustion. Failed query--environment combinations are identified explicitly and excluded rather than replaced by estimates.

The query-driven cardinality estimator is deliberately lightweight, and a non-redundant construction covers only the joint addition of mandatory patterns. The conclusions, therefore, apply to the evaluated implementation and successful query--environment combinations, rather than to every multi-model workload or optimizer architecture.
\section{Related Work}
\label{sec:related-work}

This work builds on established results in relational, distributed, and heterogeneous query optimization but applies them at a different abstraction boundary. Predicate pushdown, dependent execution, cardinality estimation, and search-space reduction are established principles. The distinguishing setting considered here begins with operators over a unified conceptual schema rather than with source-specific relations or a physical plan in which sources have already been assigned. Query fragments must first be covered by explicit physical mappings and only then translated into relational, document, or graph query languages. An optimization decision must consequently preserve conceptual semantics while satisfying both mapping coverage and native DBMS capabilities.

\subsection{Query-Plan Construction and Estimation}

Query optimization searches among semantically equivalent plans using cost and cardinality estimates. Practical optimizers restrict plan enumeration because the number of alternatives grows rapidly with the number of inputs. Distributed and heterogeneous optimization must additionally consider data placement, source capabilities, communication, and processing of intermediate results~\citep{opt-diverse-datasrc,opt-distributed,distributed-db}.

Query-driven estimators improve predictions using information collected from previous executions. Self-tuning histograms use differences between estimated and observed cardinalities~\citep{cost-query-driven-1}; QuickSel learns multidimensional selectivity models from query feedback~\citep{cost-query-driven-2}; and sampling estimates cardinalities without processing the entire dataset. Recent multistore work applies learned cost models to plans combining native and middleware operations~\citep{multistore-cost}, while MongoDB uses trial execution and recorded plan behavior during native optimization~\citep{mongodb-optimizer}.

MM-quecat uses a narrower, task-specific mechanism. Observations from structurally similar query parts and available native estimates are used to distinguish smaller and larger inputs when orienting dependent joins. Because source-specific estimates have different meanings and scales, the method does not combine them into one global cost model across PostgreSQL, MongoDB, and Neo4j.

The planning optimization addresses a different search space. In MM-quecat, the same set of mapped kinds may be reached through several insertion orders during query-part construction. Mandatory kinds are therefore added jointly. This eliminates provably redundant permutations while preserving candidates corresponding to genuinely different physical mappings. Unlike conventional join-order reduction, the alternatives are combinations of mapping-derived kind patterns rather than only orders of relational operators.

\subsection{Predicate Pushdown and Dependent Joins}

Predicate pushdown evaluates filters close to the stored data, thereby reducing intermediate-result processing and transfer~\citep{opt-diverse-datasrc,opt-distributed}. In a heterogeneous system, however, an algebraically valid transformation is executable only when the selected source supports the required operators, data types, and structural access paths.

In the proposed method, model-independent QEP transformation is therefore separated from mapping and capability validation. A conceptual predicate is incorporated into SQL, a MongoDB aggregation pipeline, or Cypher only when the selected mapping provides all required values and the corresponding native query builder supports the complete condition. Otherwise, the predicate remains in the unifying layer.

Dependent joins evaluate one input first and use its values to restrict the other. Related methods include bind joins, D-joins, lateral joins, tuple-substitution joins, and semijoin-style reduction~\citep{opt-diverse-datasrc,opt-orthogonal,distributed-db}. They are most effective when one input is small and strongly restricts the other, but may become ineffective when the generated predicate is large or weakly selective.

The adaptation considered here differs in the level at which the dependency is represented. The restriction is created over variables of the unified QEP, propagated through a heterogeneous dependent subtree, and translated only when it reaches a native query part that provides the required variable and supports the generated condition. The dependent side is therefore not restricted to a single remote relation or a single datasource leaf; it may contain model-independent operators and several native query parts. The original non-dependent join is retained as a fallback when the restriction cannot be applied safely.

\subsection{Multi-Model Languages and Systems}

Multi-model query languages provide unified access to heterogeneous representations either by extending an established language or by introducing model-independent abstractions. Cross-model joins, result construction, interoperability, and optimization remain important open problems~\citep{mm-taming,mm-look-forward}.

SQL++ generalizes relational tuples and nested JSON values through a configurable semi-structured data model and language~\citep{sqlpp}. Multi-SQL supports relational, document, graph, and key-value data through a unified language connected to model-specific processing components~\citep{multi-sql-system}. Other systems expose selected combinations of models. ArangoDB's AQL combines document and graph processing, Neo4j uses Cypher for graph-pattern queries, and MongoDB uses aggregation pipelines for document transformations~\citep{aql-manual,cypher-manual,mongodb}. Their different operator sets and structural assumptions illustrate why a unified optimizer cannot rely on a single common native capability model.

MMQL uses a category-theory-based conceptual representation and explicit mappings from conceptual schema fragments to physical kinds~\citep{mmcat,mmquecat}. A user query does not directly select a data model, DBMS, or native query language. The processor must first identify mappings that cover the conceptual query, and only then construct native query parts. The present work does not introduce another multi-model language; it addresses optimization within this mapping-driven decomposition process.

Multistore and polystore systems similarly coordinate specialized engines. Their optimizers consider engine assignment, data movement, conversion, and middleware processing, and recent work applies learned models to plans combining native and middleware operators~\citep{multistore-cost}. These systems confirm that heterogeneous optimization cannot be reduced to native execution cost alone.

In the architecture considered here, physical query components are derived from explicit mappings over a single conceptual schema. Predicate pushdown and dependent execution are introduced before native translation but remain constrained by mapping coverage and target capabilities. Non-redun\-dant construction operates over combinations of mappings rather than only over conventional physical operators or engine assignments.

\subsection{Evaluation and Positioning}

UniBench evaluates mixed-model workloads spanning relational, document, graph, key-value, and XML data~\citep{unibench-journal}, while PolyBench focuses on cross-engine execution in polystore systems~\citep{polybench}. Both demonstrate that heterogeneous performance depends on data placement, engine assignment, transfer, and coordination in addition to native execution.

The evaluation in this paper has a narrower purpose. It neither introduces a general-purpose benchmark nor compares complete multi-model systems. It examines three optimization mechanisms within one decomposition-based processor and separately measures planning, combined native execution, and result retrieval, intermediate-result processing, and final result construction.

The positioning of the work is defined by the combination of three properties. First, predicates move from a conceptual QEP into native relational, document, and graph queries only after mapping and capability validation. Second, dependent restrictions are represented at the model-independent plan level and may propagate through heterogeneous subtrees before native translation. Third, mandatory mapping-derived kind patterns are added without exploring equivalent construction orders. Together, these mechanisms span conceptual semantics, physical mappings, data-model-specific translation, and heterogeneous execution capabilities.
\section{Conclusion}
\label{sec:conclusion}

Heterogeneous query processing incurs costs not only because data are stored in different models and systems, but also because a unifying layer may make avoidable decisions about operator placement, intermediate-result retrieval, and query-part construction. We refer to these additional costs in the evaluated decomposition-based architecture as the \emph{cross-model tax}.

This paper addressed three sources of such overhead within a mapping- and capability-aware optimization pipeline. Model-aware predicate pushdown moves applicable filters into native PostgreSQL, MongoDB, and Neo4j queries. Cross-model dependent joins use values produced by one plan branch to restrict another before its data are retrieved. Non-redundant query-part construction adds mandatory kind patterns without exploring equivalent insertion orders. A lightweight query-driven cardinality estimator supports dependent-join orientation without attempting to replace the heterogeneous cost models of the underlying systems.

The implementation in MM-quecat demonstrates these adaptations over relational, document, graph, and heterogeneous environments. For individual query--environment combinations, the maximum observed latency reduction after predicate pushdown reached two orders of magnitude, and the optimization prevented the out-of-memory failures observed in the original single-DBMS experiments. Eligible dependent joins provided further improvements of up to one order of magnitude. For the largest evaluated Neo4j plans, non-redundant construction reduced planning time from almost 600 milliseconds to several milliseconds. These maxima characterize particular combinations rather than uniform improvements across the workload.

The results support a consistent design principle for decomposition-based multi-model processors: operations should cross a native query boundary only when they cannot be delegated safely. In the evaluated architecture and workload, moving filters and dependent restrictions into native query components reduced the number of results retrieved, converted, and processed by the unifying layer. Applying established optimization principles in this setting nevertheless requires more than algebraic rewriting: conceptual semantics, physical mapping coverage, data-model-specific representation, and native execution capabilities must be considered jointly.

The study is limited to inner select--project--join queries with filters, synthetically generated transactional data, and a prototype deployment in which all systems run in containers on one physical machine. The cardinality estimator is deliberately task-specific, and the planning optimization eliminates only permutations of mandatory kind patterns. Future work includes global cost-based plan and mapping selection, broader join ordering, and pushdown of projections, aggregations, sorting, and result-size restrictions. These limitations bound the empirical conclusions to the evaluated architecture and workload, while the proposed pipeline identifies the interfaces at which further multi-model optimization methods can be incorporated.

\backmatter

\section*{Declarations}

\subsection*{Funding}
This work was supported in part by the Grant Agency of the Czech Republic (GA\v{C}R), grant no. 23-07781S.

\subsection*{Competing Interests}
The authors have no relevant financial or non-financial interests to disclose.

\subsection*{Data Availability}
The synthetic dataset and experimental artifacts used in this study are publicly available in Zenodo at \url{https://doi.org/10.5281/zenodo.22675597}. The deposit contains procedurally generated data, based on a schema inspired by the Cal.com scheduling application, in three equivalent representations for PostgreSQL, MongoDB, and Neo4j, together with the supporting data-generation code and raw performance-evaluation outputs.

\subsection*{Code Availability}
The implementation of the optimization methods evaluated in this study is available at \url{https://github.com/mmcatdb/mmcat/tree/cross-model-tax-v1.0}.

\subsection*{Author Contributions}
Filip \v{S}trobl developed the proposed methods, implemented the optimizations, conducted the experiments, and prepared the initial manuscript. J\'achym B\'art\'ik contributed to the development of MM-quecat and to the analysis and interpretation of the results. Irena Holubov\'a supervised the research, contributed to the conceptualization and methodology, and revised the manuscript. All authors read and approved the final manuscript.

\bibliography{references}

\end{document}